\documentclass{article}
\usepackage{authblk}
\usepackage[utf8]{inputenc}
\usepackage[round]{natbib}
\usepackage[pdftex]{graphicx}
\usepackage{amsmath,amsthm,amsfonts}
\usepackage{lscape}   
\usepackage{rotating}
\usepackage{rotfloat}
\usepackage{amsthm}
\usepackage{multirow}
\usepackage{booktabs}
\usepackage{ctable}
\usepackage{threeparttable}
\usepackage{booktabs}
\usepackage{framed}
\usepackage{color}
\usepackage[colorlinks=true,linkcolor=blue,allcolors=blue]{hyperref}

\usepackage{soul}
\usepackage{hyperref}
\newtheorem{proposition}{Proposition}

\title{Bayesian wavelet shrinkage for low SNR data based on the Epanechnikov kernel}

\author{Fidel Aniano Causil Barrios}
\author{Alex Rodrigo dos S. Sousa}

\affil{University of Campinas (UNICAMP)\\ Departament of Statistics, Brazil \thanks{Barrios (f244960@dac.unicamp.br) and Sousa (asousa@unicamp.br)}}

\date{} 

\begin{document}

\maketitle

\vspace{-1.0cm}
\begin{abstract}
Consider the univariate nonparametric regression model with additive Gaussian noise and the representation of the unknown regression function in terms of a wavelet basis. We propose a shrinkage rule to estimate the wavelet coefficients obtained by mixing a point mass function at zero with the Epanechnikov distribution as a prior for the coefficients. The proposed rule proved to be suitable for application in scenarios with low signal-to-noise ratio datasets and outperformed standard and Bayesian methods in simulation studies. Statistical properties, such as squared bias and variance, are provided, and an explicit expression of the rule is obtained. An application of the rule is demonstrated using a real EEG dataset.

  \end{abstract}

  

\section{Introduction}
Wavelet-based methods have proven useful in several fields, including signal processing, image compression, and statistical analysis. In statistics, notable applications include nonparametric regression, density estimation, functional data analysis, and time series analysis. In nonparametric regression, the focus of this work, the representation of an unknown function as a linear combination of wavelet basis is well localized in both time and scale domains, which is an advantage over other basis, such as splines (and their variants) and the Fourier, for instance. Additionally, the wavelet representation is usually sparse, i.e., most of the coefficients in the expansion are zero, allowing the main features of the function, such as peaks, discontinuities, and oscillations, to be represented in a few significant coefficients and through multiscale visualizations. See \cite{daubechies1992} and \cite{mallat2009} for a theoretical overview of wavelets, but also \cite{vidakovic1999} and \cite{nason2008} for applications of wavelets in statistics.

The estimation of the wavelet coefficients takes into account the sparsity property of these coefficients. The estimators typically shrink the empirical (observed) wavelet coefficients obtained by applying a discrete wavelet transform (DWT) on the original data in such a way that small empirical coefficients are shrunk toward zero. Standard estimators are thresholding rules, i.e., they estimate a wavelet coefficient as zero if its associated empirical coefficient is less than a threshold value. See \cite{dj1994, dj1995} for the seminal works in wavelet shrinkage with the proposal of the so-called Donoho and Johnstone soft and hard thresholding rules. 

Bayesian approaches have also been extensively explored because they allow the incorporation of prior information, such as the sparsity property and the support of the coefficients, which is helpful in improving the estimates. 
The standard prior to the wavelet coefficients is of spike and slab type. \cite{chipman1997} proposed a mixture of two normal distributions, one of them extremely concentrated around zero (the spike) and the other with a variance that allows nonzero coefficients (the slab). Other proposals include a mixture of a point mass function at zero (the spike) and a symmetric and unimodal distribution (the slab), such as the double exponential, the uniform, the double Weibull, the beta and the logistic distributions proposed by \cite{vidakovic-ruggeri-bams}, \cite{angelini-vidakovic-2004}, \cite{remenyi2015}, \cite{alex-beta} and \cite{alex-logistica} respectively.  

Although the shrinkage rules available in the literature have been successfully applied to several real data problems, they often do not perform well in data with high noise levels. Recently, \cite{vimalajeewa2023} proposed a Bayesian shrinkage rule that addresses this question, but it is still necessary to carry out further studies of shrinkage rules that can be applied in this context of data with low signal-to-noise ratio (SNR). In this sense, we propose a Bayesian shrinkage rule based on a mixture of a point mass function at zero and the Epanechnikov distribution as a prior distribution for the wavelet coefficients. The proposed shrinkage rule performs well in low-SNR datasets and outperforms standard and Bayesian methods in general simulation studies. Furthermore, the hyperparameters of the proposed prior are closely related to the degree of shrinkage imposed by the rule, allowing us to elicit them according to the required smoothness in the data.

This paper is organized as follows: Section 2 defines the nonparametric regression model in the time and wavelet domains and the proposed prior distribution of the wavelet coefficients based on the Epanechnikov distribution. The associated Bayesian shrinkage rule and its statistical properties, such as squared bias, variance, and risks, are developed in Section 3. Simulation studies to evaluate the performance of the proposed rule and to compare it with standard methods are analyzed in Section 4. Section 5 provides a real data illustration involving an electroencephalogram (EEG) dataset.  Final considerations and further remarks are in Section 6.
\section{Statistical model}

Consider $n = 2^J$, with $J\in \mathbb{N}$, observations $(x_1,y_1),\ldots, (x_n,y_n)$, and the nonparametric regression model
\begin{equation}
    \label{time_model}
    y_i= f(x_i) +\epsilon_i, \hspace{0.5cm} i=1,\ldots,n
\end{equation}
where $x_1,...,x_n$ are equally spaced scalars, $f\in \mathbb{L}_2 \left(\mathbb{R} \right) = \{f: \int f^2(x)dx < \infty\}$ is an unknown squared integrable function and $\epsilon_i$ are independent and identically distributed (IID) normal random errors with zero mean and unknown variance $\sigma^2$, $\sigma > 0$, i.e. $\epsilon_i\sim \mathrm{N}(0,\sigma^2)$. The goal is to estimate the function $f$ based on the observations without making any assumptions about its functional structure. The standard nonparametric procedure is to represent $f$ as a linear combination of basis functions for  $\mathbb{L}_2(\mathbb{R})$ and estimate the coefficients of the expansion. Several basis functions are candidates for expanding $f$, such as polynomials, splines and their variations, the Fourier basis, and wavelets; see \cite{takezawa-2005} for more details. We consider in this work the expansion of $f$ in a wavelet basis,
\begin{equation} \label{expan}
f(x) = \sum_{j,k \in \mathbb{Z}}\theta_{j,k} \psi_{j,k}(x), 
\end{equation}
where $\{\psi_{j,k}(x) = 2^{j/2} \psi(2^j x - k),j,k \in \mathbb{Z} \}$ is an orthonormal wavelet basis for $\mathbb{L}_2(\mathbb{R})$ constructed by dilations $j$ and translates $k$ of a function $\psi$ called a wavelet or mother wavelet and $\theta_{j,k} \in \mathbb{R}$ are wavelet coefficients that describe features of $f$ at spatial locations $2^{-j}k$ and scales $2^j$ or resolution levels $j$. Thus, according to the representation \eqref{expan}, the problem of estimating the function $f$ is reduced to the problem of estimating the wavelet coefficients $\theta_{j,k}$. For simplicity, the subscripts $ j$ and $ k$ will be dropped in the text without loss of interpretation.

In order to estimate the wavelet coefficients, we take the original observations in the time domain to the wavelet domain by the application of the DWT. The model \eqref{time_model} can be written in vector notation as
\begin{equation}
    \label{vector_model}
    \boldsymbol{y}=\boldsymbol{f}+\boldsymbol{\epsilon},
\end{equation}
where $\boldsymbol{y} = [y_1,\ldots, y_n]'$, $\boldsymbol{f}=[f(x_1),\ldots,f(x_n)]'$ and $\boldsymbol{\epsilon}=[\epsilon_1,\ldots,\epsilon_n]'$. The DWT can be represented by a $n \times n$ orthogonal transformation matrix $\boldsymbol{W}$, which is applied on both sides of \eqref{vector_model}, obtaining the following model in the wavelet domain
\begin{equation}
    \label{wavelet_model}
    \boldsymbol{d}=\boldsymbol{\theta}+\boldsymbol{\varepsilon},
\end{equation}
where $\boldsymbol{d}=\boldsymbol{Wy} = [d_1,\ldots,d_n]'$ is the vector of empirical wavelet coefficients, $\boldsymbol{\theta}=\boldsymbol{Wf} = [\theta_1,\ldots,\theta_n]'$ is the vector of wavelet coefficients and $\boldsymbol{\varepsilon}=\boldsymbol{W\epsilon} = [\varepsilon_1,\ldots,\varepsilon_n]'$ is the vector of random errors. Thus, we can interpret a specific empirical wavelet coefficient $d$ in \eqref{wavelet_model} as a noisy version of its associated wavelet coefficient $\theta$ by the random error $\varepsilon$. Further, since $\boldsymbol{W}$ is orthogonal, the random errors in the wavelet domain $\varepsilon_i$ remain IID normally distributed with zero mean and variance $\sigma^2$. 

After the application of the DWT, we apply a shrinkage rule $\delta(\cdot)$ on the empirical coefficients to estimate the wavelet coefficients, i.e, $\hat{\theta} = \delta(d)$, where $\hat{\theta}$ is the estimate of $\theta$. The shrinkage rule typically operates by reducing the magnitude of the empirical coefficient to estimate the wavelet coefficient. 

\begin{figure}
    \centering
    \includegraphics[width=1\linewidth]{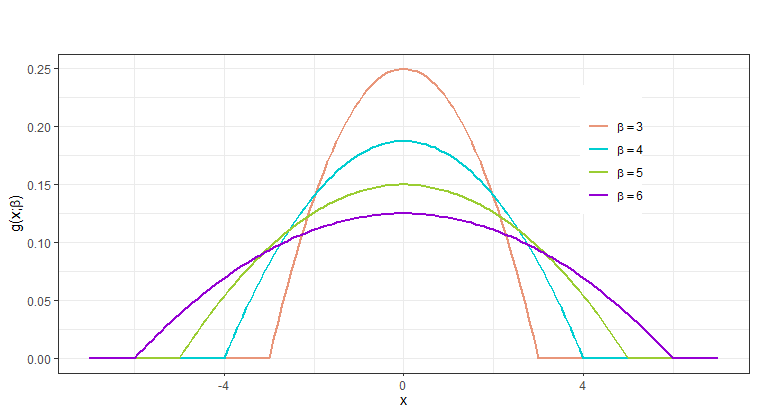} 
    \caption{Epanechnikov density function \eqref{epaneck} for $\beta \in \{3,4,5,6\}$.}
    \label{edf:app}
\end{figure}

Under a Bayesian perspective, a common prior distribution $\pi(\cdot)$ to the wavelet coefficient is the mixture of a point mass function at zero $\delta_0(\cdot)$ and a symmetric, around zero, unimodal density $g(\cdot)$, 
\begin{equation}
    \label{prior}
    \pi(\theta;\alpha)=\alpha \delta_0(\theta)+(1-\alpha)g(\theta),
\end{equation}
where $\alpha\in(0,1)$ is a hyperparameter. In this work, we propose $g(\cdot)$ to be 
\begin{equation}
\label{epaneck}
g(\theta) = g(\theta;\beta)=\frac{3}{4\beta^3}\left(\beta^2 -\theta^2\right)\mathbb{I}_{(-\beta,\beta)}(\theta),
\end{equation}
where $\beta > 0$ and $\mathbb{I}_A(\cdot)$ is the indicator function on the set $A$. Thus, the proposed prior distribution for the wavelet coefficient $\theta$ has hyperparameters $\alpha$ and $\beta$ that need to be elicited. Moreover, these hyperparameters have an impact on the severity of the shrinkage rule, which reduces the magnitude of the empirical coefficient. The proposed prior distribution \eqref{prior} and \eqref{epaneck} is based on the Epanechnikov kernel function
\begin{equation}
\label{Kernel_Epa}
K(z)= \frac{3}{4}(1 - z^2) \mathbb{I}_{(-1,1)}(z), \nonumber
\end{equation}
which is widely applied in statistics due to its efficiency and ability to effectively smooth data locally, as well as its low mean squared error in nonparametric estimation. See \cite{silverman-1986} for more details. For this reason, we will call \eqref{epaneck} the Epanechnikov density function (EDF). Figure \ref{edf:app} presents the EDF for $\beta \in \{3,4,5,6\}$. We observe that the density becomes more concentrated around zero when $\beta$ decreases. This characteristic will be important in determining the value of $\beta$ in the shrinkage rule, according to the desired level of shrinkage to be imposed on the empirical coefficients. 

\section{Bayesian shrinkage rule and its properties}

The shrinkage rule $\delta(\cdot)$ associated with the model \eqref{wavelet_model} and prior distribution \eqref{prior} and \eqref{epaneck} is obtained under the quadratic loss function $L(\delta, \theta) = (\delta - \theta)^2$. In this case, it is given by the posterior mean of $\theta$, i.e, $\delta(d) = \mathbb{E}(\theta|d)$. Proposition \ref{prop_rule} presents the explicit formula of the proposed Bayesian shrinkage rule, which we will call the Epanechnikov shrinkage rule (ESR). The proof of the proposition is in the Appendix. 

\begin{proposition}
 \label{prop_rule}

Consider the model \eqref{wavelet_model} and the prior distribution \eqref{prior} and \eqref{epaneck} to the wavelet coefficient $\theta$. In addition, assume an exponential distribution with parameter \(\lambda > 0\) as a prior distribution to \(\sigma^2\). Then, the Bayesian shrinkage rule $\delta(d)$ obtained under the quadratic loss function is expressed by
\begin{equation}
\label{rule} 
\delta(d)=\frac{(1-\alpha) \frac{3\sqrt{ 2 \lambda } }{ 8\beta^3 }\left[ \frac{2\lambda\beta^2 +3\sqrt{2\lambda}\beta+3}{2\lambda^2}\left(e^{-\sqrt{2\lambda}(\beta-d)}- e^{-\sqrt{2\lambda}(\beta+d)}\right)+ \frac{(\lambda\beta^2 -3)\sqrt{2\lambda}d -\lambda\sqrt{2\lambda}d^3}{\lambda^2}\right]}{\alpha \mathcal{ED}\left(0,\frac{1}{\sqrt{2\lambda}}\right)+(1-\alpha) \frac{3\sqrt{2\lambda}}{8\beta^3}\left[ \frac{\beta}{\lambda}\left( e^{-\sqrt{2\lambda}(\beta+d)} + e^{-\sqrt{2\lambda}(\beta-d)} \right)+ \frac{2}{\sqrt{2\lambda}}\left( \beta^2 - d^2 -\frac{1}{\lambda}\right)\right]},
\end{equation}

\noindent where \( \mathcal{ED}\left(0, \frac{1}{\sqrt{2\lambda}}\right) \) is the probability density function of the double exponential distribution with mean equals to zero and scale parameter equals to \( \frac{1}{\sqrt{2\lambda}} \).

\end{proposition}

\begin{figure}
    \centering
    \includegraphics[width=.8\linewidth]{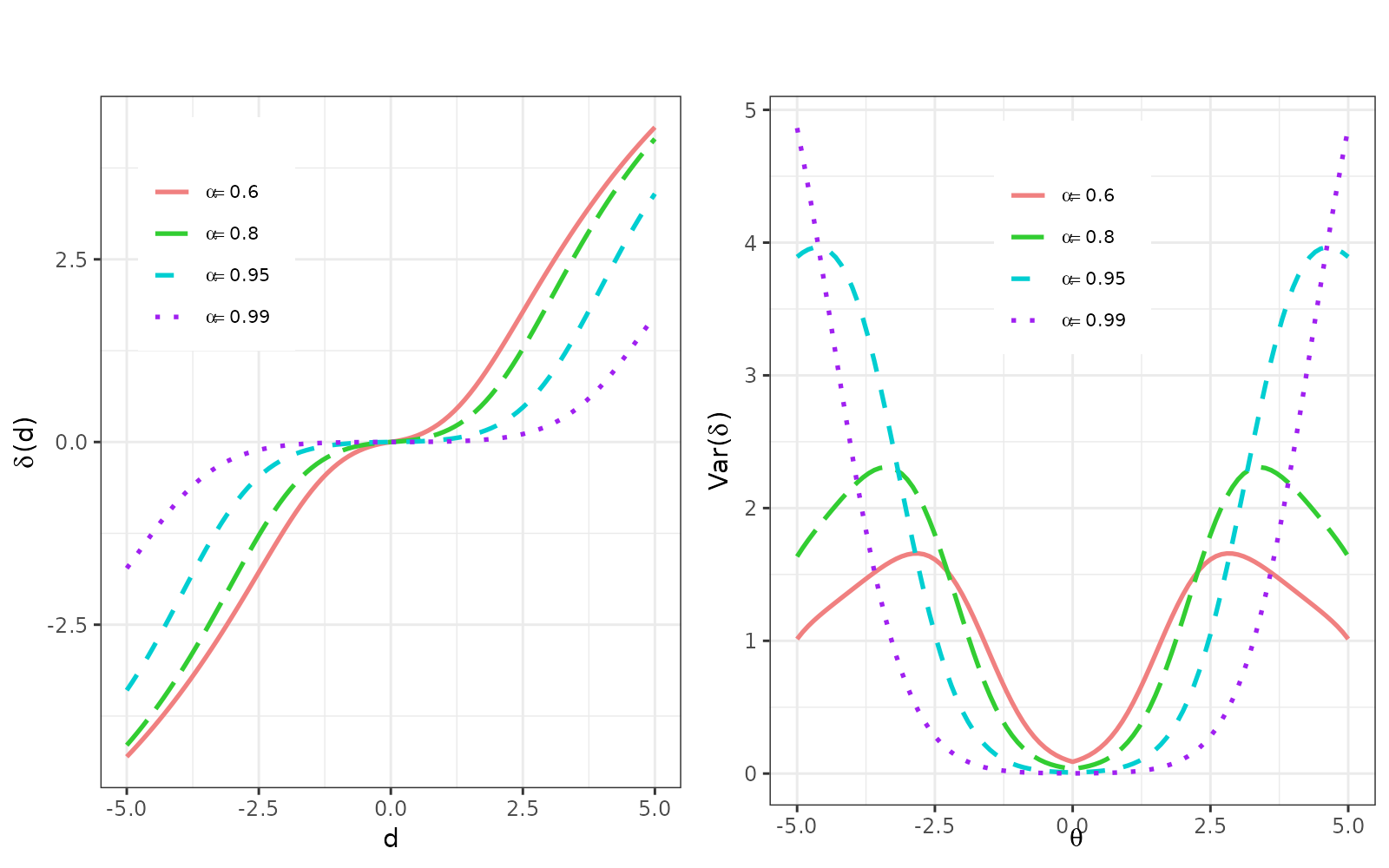} 
    \caption{Epanechnikov shrinkage rule \eqref{rule} for $\beta = 6$, $\lambda = 3$ and $\alpha \in \{0.6, 0.8, 0.95, 0.99 \}$ (left) and its associated variance (right).}
    \label{rulevar_alpha}
\end{figure}

\begin{figure}
    \centering
    \includegraphics[width=.8\linewidth]{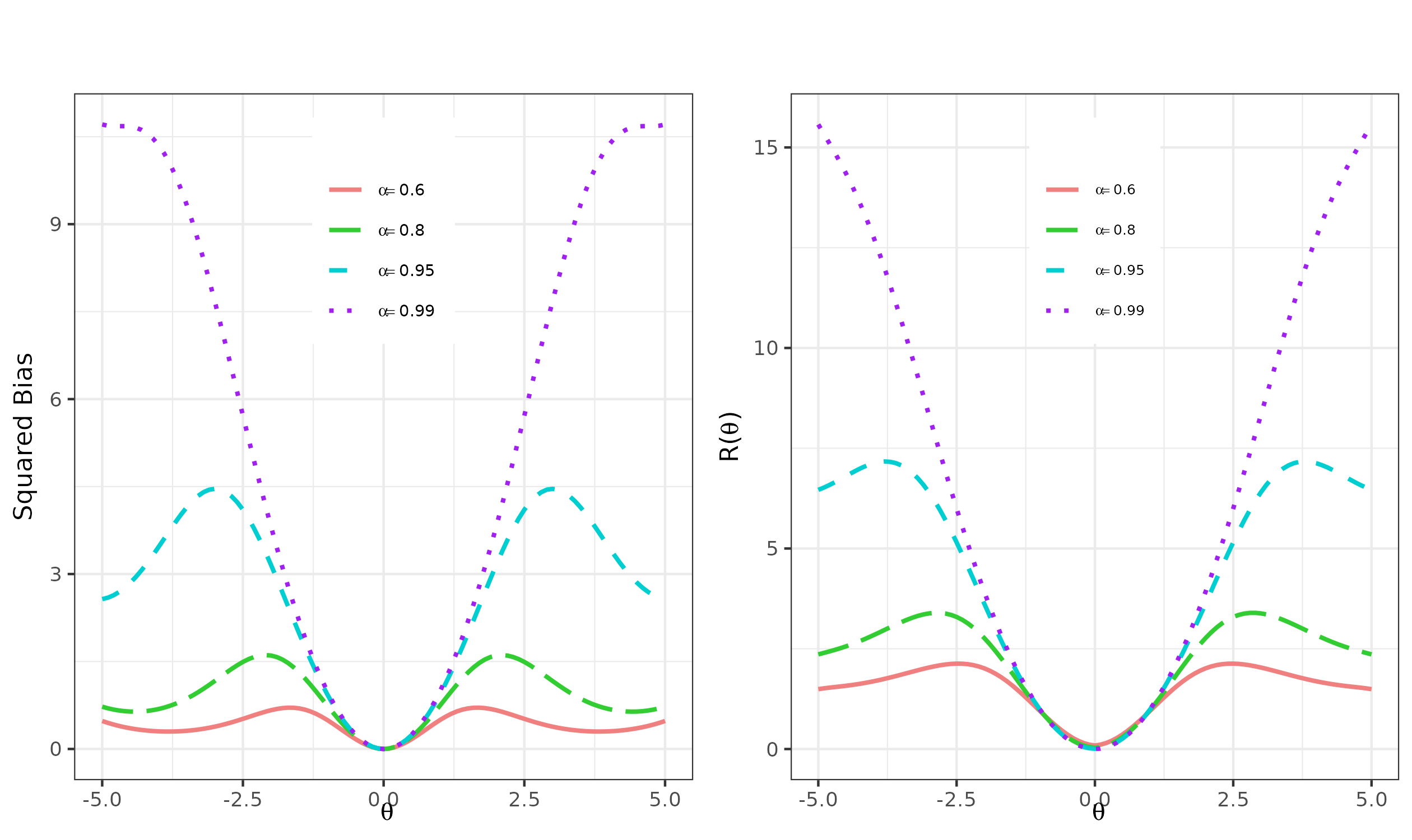} 
    \caption{Squared bias (left) and classical risk (right) for the Epanechnikov shrinkage rule \eqref{rule} for $\beta = 6$, $\lambda = 3$ and $\alpha \in \{0.6, 0.8, 0.95, 0.99 \}$.}
    \label{biasrisk_alpha}
\end{figure}

Figure \ref{rulevar_alpha} (left) illustrates the Epanechnikov shrinkage rule \eqref{rule} using the hyperparameters $\beta = 6$, $\lambda = 3$, and $\alpha \in \{0.6, 0.8, 0.95, 0.99\}$. It can be observed that the interval of $d$ where the rule reduces to zero expands as the hyperparameter $\alpha$ increases, which is reasonable since the weight of the point mass function at zero in \eqref{prior} is bigger. Furthermore, $\delta(d)$ approaches $\beta$ as $d$ increases, while for decreasing values of $d$, $\delta(d)$ approaches $-\beta$, that is, $-\beta\leq \delta(d)\leq \beta$, and empirical coefficient values outside this interval are attributed to the presence of noise. Figure \ref{rulevar_alpha} (right) presents the variance of the same shrinkage rules as those in the left figure. We observe that the variance is generally small for values of $\theta$ close to zero and then increases symmetrically until it reaches a peak. After that, it decreases until it reaches a stable level. This peak occurs at the point where the rule stops shrinking the empirical coefficients to zero. Note also that in the region near zero, the variance is lower for higher values of $\alpha$. However, beyond the peaks, this order reverses, and smaller $\alpha$ values start to exhibit lower variance. Still for the same rules, we present the squared bias and the classical risk $R(\theta) = \mathbb{E}^{d|\theta}[L(\delta(d),\theta)]$ in Figure \ref{biasrisk_alpha} (left and right respectively), which showed similar behavior. Both the bias and the risk increase up to a certain point and then decrease, as observed for the variance. However, bias and risk are higher for larger values of $\alpha$.

\begin{figure}
    \centering
    \includegraphics[width=.8\linewidth]{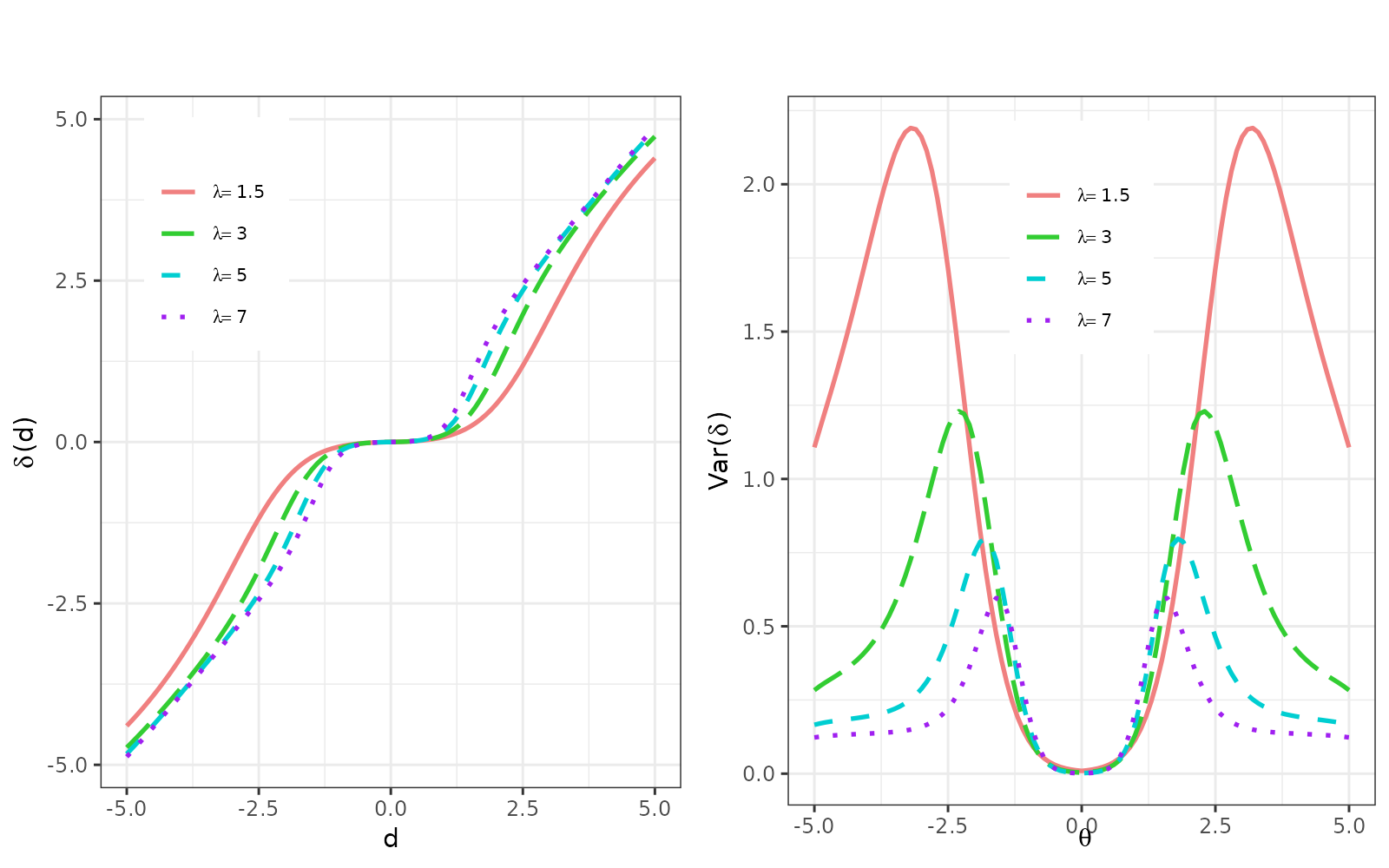} 
    \caption{Epanechnikov shrinkage rule \eqref{rule} for $\beta = 6$, $\alpha = 0.95$ and $\lambda \in \{1.5, 3, 5, 7 \}$ (left) and its associated variance (right).}
    \label{rulevar_lambda}
\end{figure}

\begin{figure}
    \centering
    \includegraphics[width=.8\linewidth]{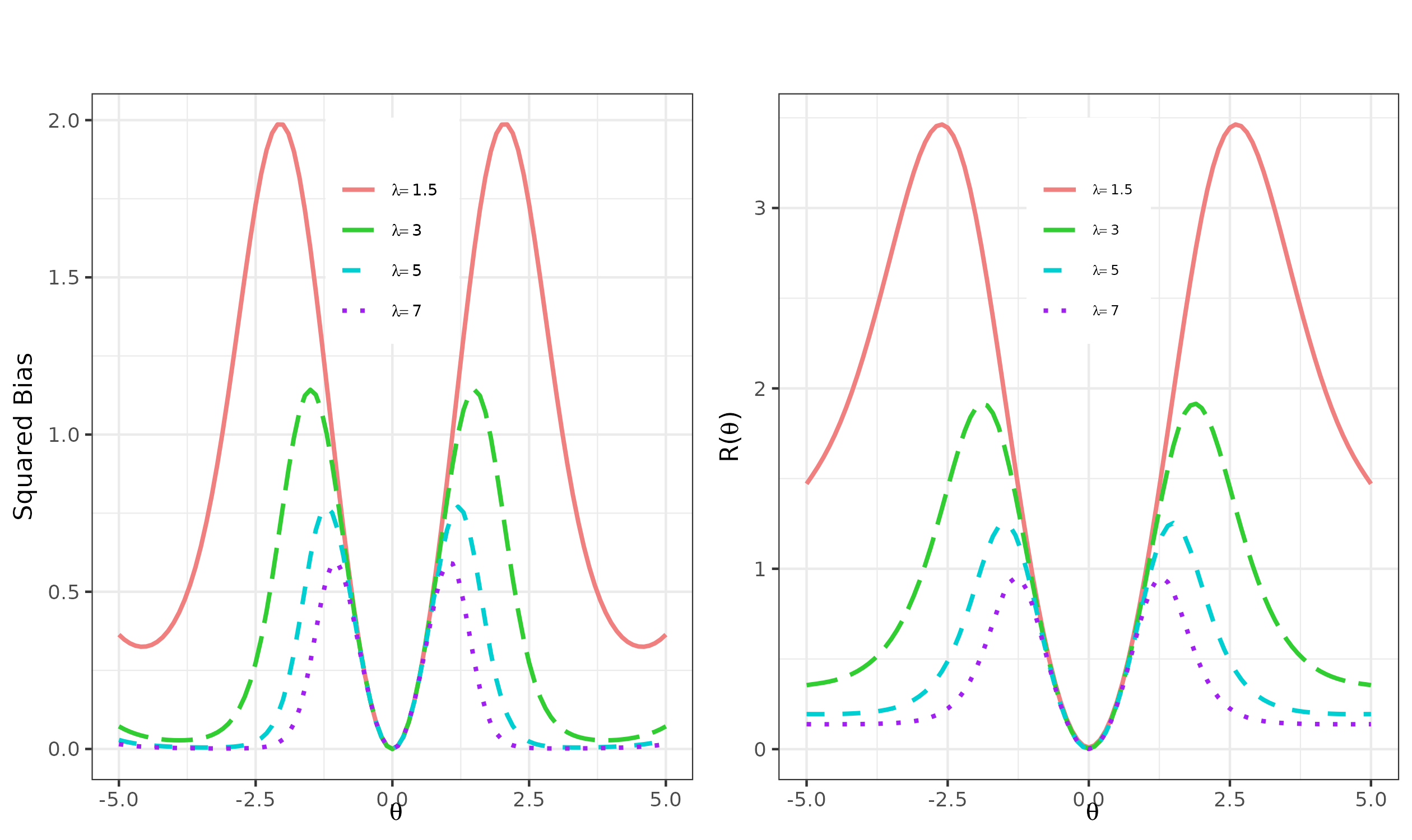} 
    \caption{Squared bias (left) and classical risk (right) for the Epanechnikov shrinkage rule \eqref{rule} for $\beta = 6$, $\alpha = 0.95$ and $\lambda \in \{1.5, 3, 5, 7 \}$.}
    \label{biasrisk_lambda}
\end{figure}

Figure \ref{rulevar_lambda} shows the Epanechnikov shrinkage rule \eqref{rule} for $\beta = 6$, $\alpha = 0.95$ and $\lambda \in \{1.5, 3, 5, 7 \}$ (left) and its associated variance (right) to evaluate the impact of the parameter $\lambda$ on the rule and its statistical properties. The general behaviors of both the shrinkage rule and variance are quite similar to those observed in Figure \ref{rulevar_alpha}, but now, when $\alpha$ is fixed, the shrinkage rule is more severe for smaller values of $\lambda$. The squared bias and classical risk behaviors displayed in Figure \ref{biasrisk_lambda} (left and right, respectively) are also similar to those in Figure \ref{biasrisk_alpha}. However, it is observed that the bias and the risk of the rule are higher for smaller values of $\lambda$.

\begin{figure}
    \centering
    \includegraphics[width=0.75\linewidth]{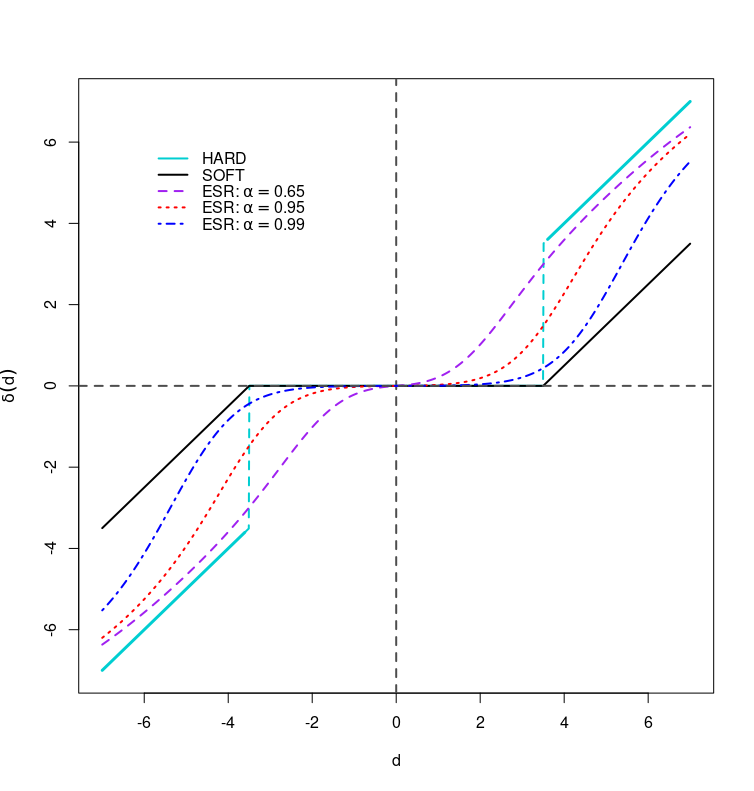} 
    \caption{ESR \eqref{rule} with $\beta=8$, $\lambda=1$, $\alpha \in \{0.65, 0.95$ and $0.99\}$ and  the hard \eqref{hardrule} and soft \eqref{softrule} thresholding rules with $\eta = 3.5$.}
    \label{rulesalpha}
\end{figure}

Finally, we compare in Figure \ref{rulesalpha} the behavior of the ESR with the so-called hard and soft thresholding rules proposed by \cite{dj1994} and given respectively by
\begin{equation}\label{hardrule} 
\delta_{\mathrm{hard}}(d) = \begin{cases} 
 0, & \text{if $|d| \leq \eta$} \\  
 d, & \text{if $|d|>\eta$,}   
 \end{cases} 
\end{equation}

\begin{equation}\label{softrule} 
\delta_{\mathrm{soft}}(d) = \begin{cases} 
 0, & \text{if $|d| \leq \eta$} \\  
 \mathrm{sgn}(d)(|d| - \eta), & \text{if $|d|>\eta$,}   
 \end{cases} 
\end{equation}
where $\eta > 0$ is the threshold and $\mathrm{sgn}(\cdot)$ is the sign function. In the figure, $\eta = 3.5$ and the hyperparameters of the ESR are $\beta = 8$, $\lambda = 1$ and $\alpha \in \{0.65,0.95,0.99\}$. Note that the ESR has an intermediate shape between the hard and soft thresholding rules according to the choice of $\alpha$. For $\alpha = 0.65$, the rule is closer to the hard thresholding rule whereas for $\alpha = 0.99$, it is closer to the soft rule. For high values of $d$, all the ESR behave as the hard thresholding rule, since the effect of the noise is irrelevant for significant empirical coefficients, which gives $\delta(d) \approx d$.

\section{Hyperparameters elicitation}

The appropriate determination of the hyperparameters used in the prior distribution \eqref{prior} and \eqref{epaneck} plays a crucial role in Bayesian methods, as it directly impacts the quality of the estimates. In this study, we use the hyperparameters $\alpha$ and $\beta$ according to the established methods and criteria discussed in the literature. The hyperparameter $\alpha$ is calculated according to the level-dependent proposal by \cite{vimalajeewa2023} given by
\begin{equation}\label{alpha}
    \alpha=\alpha(j)=1-\frac{1}{(j-J_0+l)^\gamma},
\end{equation}
\noindent where \(J_0\leq j\leq J-1\), $J_0$ is the primary resolution level, \(\gamma\) and \(l\) are positive constants. The authors suggest \(l=2\) and \(\gamma=2.4\) if no other information is available. Note that $\alpha$ increases as the resolution level $j$ increases, which means that the proposed shrinkage rule acts more severely on the empirical coefficients of the highest resolution levels. Table \ref{tab:alphas} presents the values of $\alpha(j)$ for $n = 1024 = 2^{10}$ and $J_0 = 5$, which we see, for instance, that $\alpha(5) = 0.8105$ but $\alpha(9) = 0.9864$, indicating that the empirical wavelet coefficients at the resolution level $j = 9$ are strongly shrunk around zero. The hyperparameter \(\beta\) is determined according to the also level-dependent approach used by \cite{alex-beta}, 
\begin{equation}\label{beta}
    \beta=\beta(j)=\underset{k}{\mathrm{max}} \{|d_{j,k}|\},
\end{equation}
\noindent where maximization is performed on the resolution level \(j\), which ranges from \(J_0\) to \(J-1\). This parameter selection is crucial in the estimation procedure, as it is associated with the bounded support of the wavelet coefficients for each resolution level, specifically $(-\beta, \beta)$.

\begin{table}[H] 
\centering
\begin{tabular}{c|ccccc}
\hline
$j$ & 5 & 6 & 7 & 8 & 9 \\ \hline \hline
 $\alpha(j)$ & 0.8105 & 0.9284 & 0.9641 & 0.9789 & 0.9864 \\ \hline  
\end{tabular}
\caption{Values of the hyperparameter $\alpha$ of the prior distribution \eqref{prior} as function of the resolution level $j$ according to the proposal \eqref{alpha} for $n = 1024$ and $J_0 = 5$.}
\label{tab:alphas}
\end{table}

\begin{figure}
    \centering
    \includegraphics[width=0.6\linewidth]{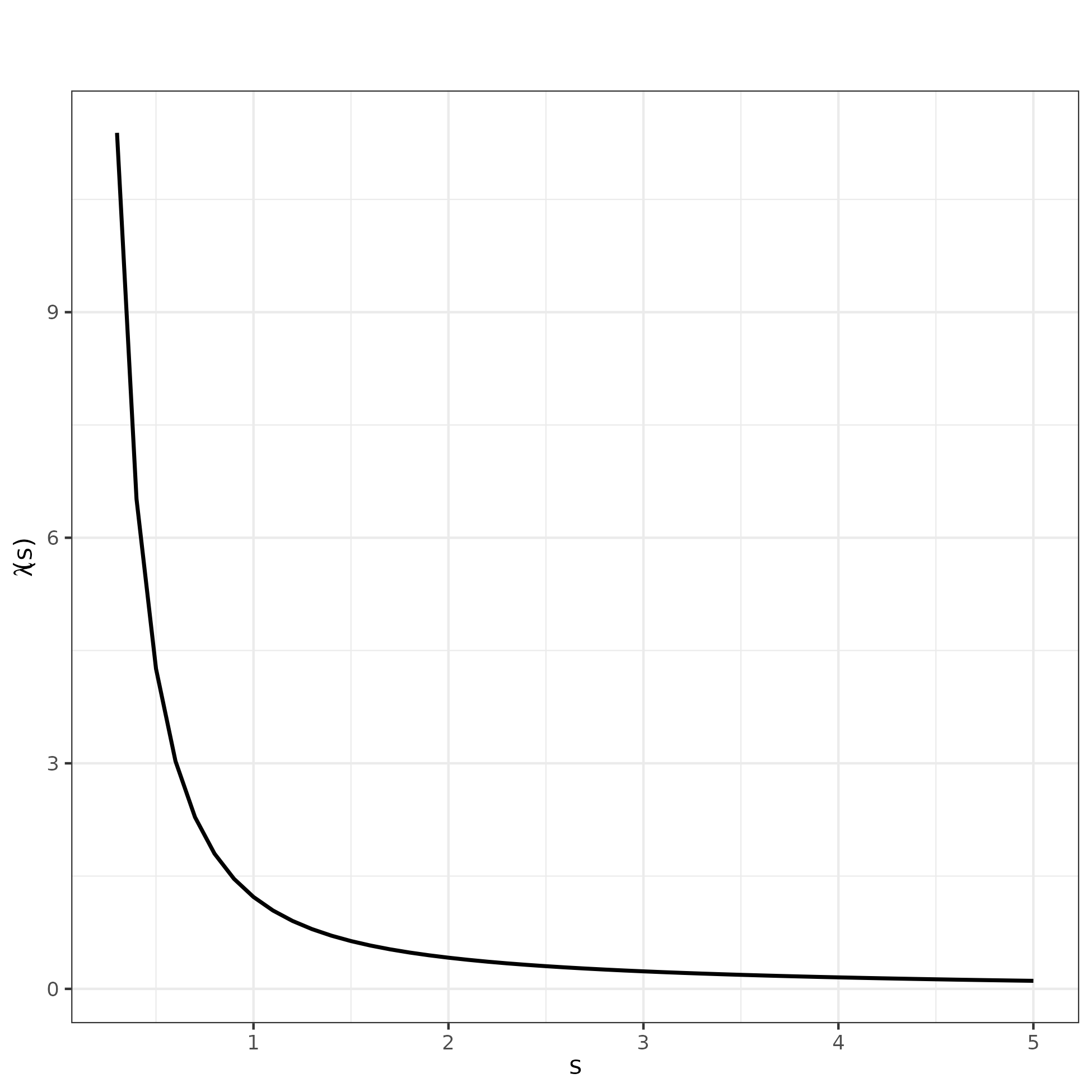} 
    \caption{The hyperparameter $\lambda$ of \eqref{rule} as a function of the sample standard deviation $s$ of the empirical wavelet coefficients at the finest resolution level according to \eqref{lambda}.}
    \label{lambda_s}
\end{figure}

Let $s=\mathrm{SD}(d_{J-1,k}: k=0,1,...,2^{J-1})$, where $\mathrm{SD}( \cdot)$ represents the usual sample standard deviation of the coefficients at the finest resolution level, which is an estimator of $\sigma$. Based on experiments to analyze the behavior of $\lambda$ as a function of $s$, we propose the following empirical expression to obtain $\lambda$,
\begin{equation}
\label{lambda}
   \lambda=\lambda(s)=\frac{1}{s^2}+\frac{c}{\tau} \exp\left(-\frac{1}{\tau} s\right),
\end{equation}
where $c>0$ and $\tau > 0$. In the absence of prior information, we recommend adopting \(c=1\) and \(\tau=2\). The expression \eqref{lambda} suggests that for $s < 1$, the behavior of $\lambda$ is closer to $1/s^2$, whereas for $s>1$, it turns to decay exponentially. Figure \ref{lambda_s} presents $\lambda$ as a function of $s$ according to the proposal \eqref{lambda}. It is also possible to consider the robust estimator 
$\hat{\sigma} = \text{median} \{ |d_{J-1, k}|,k = 0, \dots, 2^{J-1} \}/0.6745$, proposed by \cite{dj1994} in \eqref{lambda} instead of $s$.

\section{Simulation studies}
We conducted simulation studies to compare the performance of the ESR against important wavelet shrinkage and thresholding methods available in the literature, including both classical and Bayesian approaches. Special attention was given to analyzing the performance of these methods in low signal-to-noise ratio (SNR) scenarios, where the presence of significant noise challenges the effectiveness of the estimators. 

\begin{figure}
    \centering
    \includegraphics[width=1\linewidth]{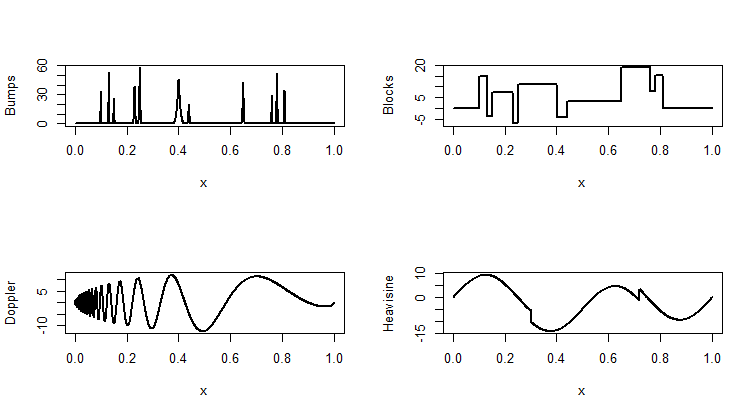} 
    \caption{Donoho and Johnstone test functions.}
    \label{djfunctions}
\end{figure}

We considered the set of four Donoho and Johnstone test functions called Bumps, Blocks, Doppler, and Heavisine as underlying functions $f(x)$ in the model \eqref{time_model}, which are shown in Figure \ref{djfunctions}. These functions were proposed by \cite{dj1994} and are widely adopted in the literature as important benchmarks for evaluating and comparing wavelet shrinkage methods due to their important local features, such as discontinuities (Blocks and Heavisine), spikes (Bumps), and oscillations (Doppler).

In the experimental scenarios for each underlying function, we generated data from model \eqref{time_model} according to three sample sizes $n = 512, 1024$ and $2048$ and the variance $\sigma^2$ of the random errors was adjusted according to three SNR values, $\mathrm{SNR} = 0.2, 1$ and $3$ by the relation $\mathrm{SNR} = \mathrm{SD}(f)/\sigma$, where $\mathrm{SD}(f)$ is the standard deviation of the underlying function (signal). For the Donoho and Johnstone test functions, $\mathrm{SD}(f) = 7$.
After generating the data, we applied a DWT to it, followed by the application of shrinkage and thresholding rules on the empirical wavelet coefficients to estimate the wavelet coefficients. We considered a Daubechies basis with ten vanishing moments in the DWT. Finally, the inverse DWT (IDWT) was performed to estimate the underlying function. Furthermore, the entire simulation study was conducted in \cite{r-version}, and the wavelet procedures were implemented using the Wavethresh R package of \cite{nason2024}, with the code available upon request.

We compared the ESR with five Bayesian methods: the BAMS proposed by \cite{vidakovic-ruggeri-bams}, the $\Gamma$-Minimax shrinkage rule by \cite{angelini-vidakovic-2004}, the Large Posterior Mode (LPM) thresholding rule by \cite{cutillo2008}, the Amplitude-scale-invariant Bayes Estimation (ABE) by \cite{figueiredo2001} and the recent shrinkage rule with beta prior by \cite{alex-beta}, the shrinkage rule under logistic prior by \cite{alex-logistica} and the three-point prior (TPP) $\Gamma$-Minimax rule by \cite{vimalajeewa2023}. Moreover, we also added the soft thresholding rule \eqref{softrule} by \cite{dj1994} with four policies to obtain the threshold value: the Universal theshold by \cite{dj1994}, the False Discovery Rate (FDR) method by \cite{abra1996}, the Cross-Validation method (CV) by \cite{nason1996} and the Stein Unbiased Risk Estimator (SURE) by \cite{dj1995}. 

In the Bayesian methods, we elicited \(\alpha\) according to  \eqref{alpha} with \(l = 1\) and \(\gamma = 2\), except for the TPP rule, which was executed with the best configuration recommended by the authors, with \(l = 5.8\) and \(\gamma = 2.4\), due to its sensitivity in parameter selection. Furthermore, we adopted the hyperparameters $a = 5$ for the beta shrinkage rule of \cite{alex-beta} and $\tau = 10$ for the logistic shrinkage rule of \cite{alex-logistica}, as these values yielded the best performance in the simulation studies conducted by the respective authors. In all methods, the primary resolution level \(J_0 = 0\) was adopted, except for the three-point prior rule, where \(J_0 = 3\) was used for \( \text{SNR} = 0.2 \) and \(J_0 = 6\) for other values. In order to evaluate the performance of the methods, we calculated the mean squared error (MSE) for each method and replication $r$,
\begin{equation*}
    \mathrm{MSE}_r=\frac{1}{n}\sum_{i=1}^{n}\left[\hat{f}^{(r)}(x_i)-f(x_i)\right]^2,
\end{equation*}
\noindent where \(\hat{f}^{(r)}(x_i)\) is the estimate of the function \(f\) at point \(x_i\), \(i=1,...,n\), in the $r$-th replication. After we ran all the $R$ replications, the average mean squared error (AMSE) was calculated for each method,
\begin{equation*}
    \mathrm{AMSE}=\frac{1}{R}\sum_{r=1}^{R} \mathrm{MSE}_r,
\end{equation*}
where $R = 300$. Considering the AMSE measure, a method's good performance is associated with a small AMSE.

Tables \ref{tab:bumps}, \ref{tab:blocks}, \ref{tab:doppler}, and \ref{tab:heavi} show the AMSE and the standard deviation of the MSE for Bumps, Blocks, Doppler, and Heavisine as underlying functions, respectively. In general, the Bayesian shrinkage rules performed better than the standard thresholding rules, except for the shrinkage rule under the logistic prior and BAMS, which had poor overall performance for low SNR. On the other hand, the proposed ESR, the rule under beta prior, and the rule under TPP outperformed the other methods in almost all the scenarios.  

\begin{figure}
    \centering
    \includegraphics[width=1\linewidth]{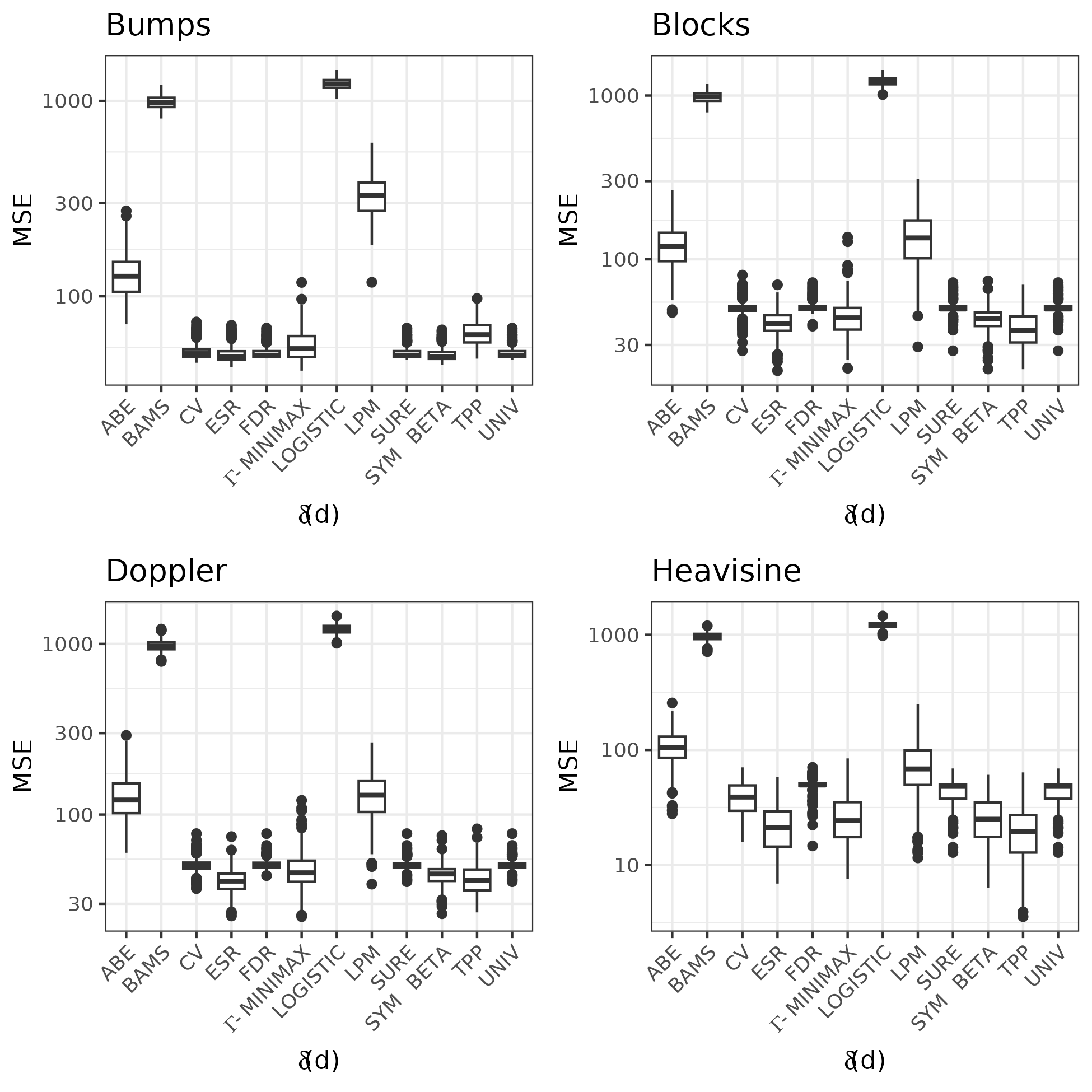} 
    \caption{Boxplots of the MSE in the simulation studies with $\mathrm{SNR} = 0.2$ and $n = 512$.}
    \label{bp}
\end{figure}

The proposed ESR had the best overall performance for the Doppler and Heavisine functions, obtaining the lowest AMSE in all scenarios of $\mathrm{SNR} = 1$ in both functions and for $\mathrm{SNR} = 0.2$ and $n=512$ and $2048$ for Doppler and $\mathrm{SNR} = 3$ and $n = 512$ and $1024$ for Heavisine. For Bumps function, it had the best AMSE when $\mathrm{SNR} = 0.2$ and $n = 1024$ and $2048$, and also for $\mathrm{SNR} = 1$ and $n = 2048$. In the other scenarios of SNR and sample size, the ESR was competitive with the best rule; for instance, when $\mathrm{SNR} = 0.2$ and $n = 512$, its AMSE was equal to $50.783$ compared to $50.622$ for the rule under a beta prior. Similar behavior was observed for the Blocks function, where the rule under TPP was the best; however, the ESR was very competitive in all scenarios, achieving the second-best AMSE in most instances. Figure \ref{bp} shows the boxplots of the MSE for $\mathrm{SNR} = 0.2$ and $n = 512$, where we see the highlighted general performance of the Bayesian rules in terms of the MSE and its small variance among the replications. The boxplots of the other scenarios of SNR and sample size are available in the Supplementary Material and show similar behavior. 

Table \ref{tab:alphajo} presents the AMSE of the ESR and the shrinkage rules under TPP and beta priors for the Blocks and Heavisine functions when $\mathrm{SNR} = 0.2$ and $n = 1024$ according to different choices of the hyperparameters $\gamma$ and $l$ in \eqref{alpha} and primary resolution levels $J_0$. We considered this evaluation to analyze the impact of these hyperparameters on the performance of the rules that showed the best overall AMSE. We considered the primary resolution level $J_0 = 0$ for all the methods in the simulation studies, except for TPP, which was run with $\gamma = 2.4$, $l = 5.8$, and primary resolution level $J_0 = 3$ for $\mathrm{SNR} = 0.2$, and $J_0 = 6$ in other cases, following the suggestion of \cite{vimalajeewa2023}. We observe that the rules exhibit similar performance for higher primary resolution levels, particularly when $J_0 = 4, 5$, and $6$. On the other hand, when we consider $J_0 = 0$ and $1$, the ESR had the smallest AMSE for both functions when $\gamma = 2$ and $l=1$, and the rule under beta prior had the smallest one when $\gamma = 2.4$ and $l = 5.8$. It suggests some robustness of the ESR and the shrinkage rule under the beta prior with relation to the choice of the primary resolution level.  


\begin{table}[H]
\centering
\begin{tabular}{|c|c|c|c|c|}
\hline
 
 \textbf{n} & \textbf{Method} & \textbf{SNR=0.2} & \textbf{SNR=1} & \textbf{SNR=3}   \\ \hline \hline
  &        & \textbf{AMSE (SD)} & \textbf{AMSE (SD)} & \textbf{AMSE (DP)} \\ \hline
&	UNIV	&	51.475 (3.723)	&	44.594 (1.571)	&	22.556 (1.547)	\\
	&	FDR	&	51.507 (3.703)	&	39.648 (3.061)	&	11.935 (1.531)		\\
	&	CV	&	52.21 (4.497)	&	24.591 (3.04)	&	9.062 (1.028)	\\
512	&SURE&	51.475 (3.723) & 44.594 (1.571)	&	4.596 (0.685)		\\

   &  ABE &  129.966 (33.821)  &   \textbf{ 21.312 (1.877) }   & 3.86 (0.432)   \\ 
	&BAMS	   &982.925 (75.825)	&	22.116 (2.19)	 &	4.162 (0.37)	\\
   & SYM BETA &   \textbf{50.622 (4.165)}    &   30.792 (3.118)  & 6.227 (0.932)       \\ 
   & LOGISTIC  &   1226.514 (79.879)   &   35.311 (13.073)   & 5.514 (0.728)        \\ 
	&	LPM	    &  332.336 (80.201)	    &	 25.085 (2.512)  &	\textbf{3.489 (0.309)}		\\
   & TPP&    65.153 (9.241) &   32.096 (1.111)     & 26.673 (0.123)        \\ 
   & $\Gamma$-MINIMAX&  57.311 (10.988)   & 25.237 (2.362)   & 5.568 (0.718)   \\ 
   &  ESR      &   50.783 (5.049)   &  26.493 (2.334)   &  8.049 (0.519) \\ \hline

 &	UNIV	&	50.041 (1.471) 	&	37.001 (1.712)	&	13.835 (0.729)	\\
	&	FDR	&	50.073 (1.451)	&	 29.581 (2.37)	&	7.045 (0.668)		\\
	&	CV	&	50.174 (1.991)	&	15.664 (1.259)	&	3.29 (0.251)	\\
1024&SURE& 50.041 (1.471) 	& 37.001 (1.712) &	3.034 (0.319)	\\

   &  ABE &   127.003 (24.253)   &   \textbf{15.085 (1.223)}   & \textbf{2.231 (0.177)}  \\ 
	&	BAMS	&962.525 (53.46)	&17.317 (1.524)	&	2.453 (0.199)		\\
   &  SYM BETA &   48.185 (2.26) &   19.014 (1.863) & 3.141 (0.345)      \\ 
   &  LOGISTIC  & 1223.709 (55.007)  &  20.492 (10.924) & 2.888 (0.297) \\ 
	&	LPM	&	315.387 (50.101)	&	20.283 (1.789)&	2.518 (0.209)\\
  & TPP &   56.254 (4.604) &    29.21 (0.565) & 26.497 (0.061)  \\ 
   &  $\Gamma$-MINIMAX&   51.887 (7.003) & 16.727 (1.329)  & 2.992 (0.307)  \\ 
   & ESR &   \textbf{47.397 (2.326)}  &   16.082 (1.298)&  3.161 (0.246)  \\ \hline

 &	UNIV	&	49.435 (1.042)	&	28.002 (1.193) &	7.836 (0.347)	\\
	&	FDR	&	49.598 (0.916)	&	21.214 (1.473)	&	4.359 (0.299)	\\
	&	CV	&	48.93 (1.596) 	&	10.944 (0.697)	&	1.861 (0.1)		\\
2048&SURE&	49.435 (1.042)	&	24.928 (7.026) & 1.984 (0.14)		\\

   &  ABE &    122.999 (17.907)   &  10.284 (0.761)     & 1.444 (0.089)     \\ 
	&	BAMS	&943.293 (34.902)	&	12.968 (1.063)	&	\textbf{1.427 (0.09)}	\\
   &  SYM BETA &   46.299 (1.99)   &   11.246 (1.008)     & 1.76 (0.127)    \\ 
   &  LOGISTIC &   1223.563 (35.467)   &  11.606 (7.462) & 1.612 (0.112)   \\ 
	&	LPM	&	305.603 (37.147)	&	16.219 (1.4)	&	2.007 (0.149)		\\
   & TPP&  51.692 (2.425)  &  27.796 (0.281)      & 26.46 (0.031)   \\ 
   & $\Gamma$-MINIMAX&   47.803 (5.012)   &  10.208 (0.824) & 1.673 (0.117) \\ 
   &  ESR  &   \textbf{45.069 (1.854)} &   \textbf{9.86 (0.772)}      & 1.654 (0.105)    \\ \hline

\end{tabular}
\caption{AMSE and standard deviation (SD) for the shrinkage and thresholding rules considered in the simulation studies. Results for Bumps as underlying function.}\label{tab:bumps}
\end{table}


\begin{table}[H]
\centering
\begin{tabular}{|c|c|c|c|c|}
 \hline
\textbf{n} & \textbf{Method} & \textbf{SNR=0.2} & \textbf{SNR=1} & \textbf{SNR=3}   \\ \hline \hline
  &        & \textbf{AMSE (SD)} & \textbf{AMSE (SD)} & \textbf{AMSE (SD)} \\ \hline
 &	UNIV	&	50.689 (4.449)	&	22.227 (1.958)	&	9.744 (0.621)	\\
	&	FDR	&	51.249 (3.897)	&	20.516 (2.626)	&	6.82 (0.755)		\\
	&	CV	&	49.739 (5.886)	&	11.759 (1.389)	&	2.918 (0.276)	\\
512	&SURE&	50.689 (4.449)	&22.225 (1.967)&	3.367 (0.515)		\\

   &  ABE     &   123.444 (35.115)    &   11.684 (1.438)    & \textbf{2.915 (0.29)}         \\ 
	&BAMS	   &978.877 (75.028)	&	16.131 (2.19)	 &	3.145 (0.24)		\\
   & SYM BETA &   43.447 (6.961)      &   12.868 (1.916)  & 4.323 (0.426)         \\ 
   & LOGISTIC &   1226.333 (78.521)   &   21.17 (23.639)   & 3.879 (0.344)         \\ 
	&	LPM	    &  140.919 (53.945)    &	 13.408 (1.855)  &	3.417 (0.342)		\\
  & TPP&   \textbf{ 38.537 (9.733) }  &   \textbf{9.518 (1.086)} & 4.024 (0.127)         \\ 
   & $\Gamma$-MINIMAX&  45.625 (13.518)   & 10.405 (1.36)   & 3.782 (0.381)  \\ 
   &  ESR     &  41.238 (7.4)   &  9.924 (1.355) & 3.515 (0.264)  \\ \hline

 &	UNIV	&	47.872 (3.55)	&	17.022 (1.339)	&	7.023 (0.341)	\\
	&	FDR	&	49.4 (2.597)	&	15.411 (1.728)	&	4.74 (0.421)		\\
	&	CV	&	44.049 (6.24)	&	9.145 (0.771)	&	2.156 (0.173)	\\
1024&SURE&	47.872 (3.55)	&16.814 (1.98)	&	2.317 (0.252)		\\

   &  ABE &   112.756 (23.74)   &   9.433 (0.972)     & 1.941 (0.158)         \\ 
	&	BAMS	&	959.629 (51.525)	&13.63 (1.585)	&	\textbf{2.072 (0.157)}		\\
   &  SYM BETA &   36.574 (6.577)   &   8.726 (0.897)      & 2.714 (0.296)        \\ 
  &  LOGISTIC  &   1227.907 (54.322)   &  11.885 (13.66)  & 2.46 (0.213)  \\ 
	&	LPM	&	124.76 (33.664)	&	10.652 (1.213)	&	2.212 (0.179)		\\
   &TPP&   \textbf{29.425 (4.834)}    &   \textbf{ 6.616 (0.496) }    & 3.886 (0.057)   \\ 
   & $\Gamma$-MINIMAX&   35.795 (8.767) & 7.601 (0.649)  & 2.384 (0.216)    \\ 
   &  ESR &   34.267 (6.149)  &  7.039 (0.636)     & 2.28 (0.167)    \\ \hline

 &	UNIV	&	43.072 (4.441)	&	13.005 (0.756)	&	4.827 (0.21)	\\
	&	FDR	&	46.913 (3.75)	&	9.869 (3.248)	&	3.293 (0.223)	\\
	&	CV	&	36.01 (5.886)	&	7.027 (0.516)	&	1.515 (0.088)		\\
2048&SURE&	43.072 (4.441)	&	9.798(3.156) & 1.615 (0.118)	\\

   &  ABE &    105.008 (17.517)   &   7.685 (0.649)      &  1.32 (0.093)    \\ 
	&	BAMS	&	939.321 (38.041)	&	11.363 (0.977)	&	\textbf{1.318 (0.086)}	\\
  &  SYM BETA &   28.271 (5.193)    &  6.329 (0.538) & 1.646 (0.128)    \\ 
   &  LOGISTIC  &   1225.802 (39.681)  &  6.692 (6.722)  & 1.529 (0.107)   \\ 
	&	LPM	&	111.069 (22.725)	&	8.715 (1.026)	&	1.459 (0.113)		\\
   & TPP &    \textbf{24.161 (2.35)}  &   \textbf{5.142 (0.286)}      & 3.771 (0.028)   \\ 
   & $\Gamma$-MINIMAX&   27.593 (6.093)   &   5.795 (0.458)  & 1.511 (0.107)   \\ 
   &  ESR &  26.369 (4.598) &   5.28 (0.432)     & 1.439 (0.095)     \\ \hline

\end{tabular}
\caption{AMSE and standard deviation (SD) for the shrinkage and thresholding rules considered in the simulation studies. Results for Blocks as underlying function.}\label{tab:blocks}
\end{table}


\begin{table}[H]
\centering
\begin{tabular}{|c|c|c|c|c|}
\hline
 
 \textbf{n} & \textbf{Method} & \textbf{SNR=0.2} & \textbf{SNR=1} & \textbf{SNR=3}   \\ \hline \hline
  &        & \textbf{AMSE (SD)} & \textbf{AMSE (SD)} & \textbf{AMSE (SD)} \\ \hline
 &	UNIV	&	50.852 (3.919) &	23.713 (2.528) &	5.943 (0.635)	\\
	&	FDR	&	51.315 (3.525)	&	20.856 (3.544)	&	4.157 (0.679)		\\
	&	CV	&	50.313 (5.408)	&	10.218 (1.367)	&	1.673 (0.208)	\\
512	&SURE&	50.852 (3.919)	&23.634 (2.829)&	1.793 (0.258)	\\

   &  ABE     &  126.958 (35.624)  &   9.579 (1.49)    & 1.283 (0.189)       \\ 
	&BAMS	   &	977.468 (71.203)	&	13.965 (2.287) &	\textbf{1.131 (0.172)}	\\
   & SYM BETA &  44.584 (6.493) &   10.48 (1.929)   &1.674 (0.282)   \\ 
   & LOGISTIC  &  1222.286 (74.967)  &   18.314 (22.73)  & 1.287 (0.212)      \\ 
	&	LPM	    &  133.789 (43.443)   &	10.685 (1.899) &	1.364 (0.249)		\\
   & TPP&   48.111 (4.438)  &   7.781 (1.087)    & 2.39 (0.116)       \\ 
   & $\Gamma$-MINIMAX&   42.615 (9.136)  & 8.139 (1.328)  & 1.32 (0.224)        \\ 
   &  ESR  & \textbf{41.295 (6.9) }  &  \textbf{7.566 (1.22)} & 1.252 (0.187)  \\ \hline

 &	UNIV	&	48.976 (2.948)	&	16.317 (1.549)	&	3.609 (0.31)	\\
	&	FDR	&	49.991 (1.933)	&	14.109 (1.998)	&	2.616 (0.343)		\\
	&	CV	&	46.518 (4.968)	&	7.042 (0.871)	&	1.165 (0.11)\\
1024&SURE&	 48.976 (2.948)	&15.243 (3.687)	&	1.22 (0.135)		\\

   &  ABE &   116.256 (23.703)   &   6.878 (0.976)      & 0.933 (0.113)        \\ 
	&	BAMS	&	957.522 (52.857)	&	11.12 (1.438)	&	\textbf{0.726 (0.083)	}	\\
   &  SYM BETA &   39.001 (5.417)   &   5.994 (0.999)      & 1.012 (0.135)   \\ 
   & LOGISTIC  &   1226.301 (54.097)  &   8.686 (15.325)     &  0.793 (0.093)      \\ 
	&	LPM	&	126.876 (33.853)	&	7.57 (1.4)	&	1.024 (0.139)		\\
   & TPP&   \textbf{33.836 (4.732)} &   4.782 (0.545)    & 2.087 (0.061)         \\ 
   & $\Gamma$-MINIMAX&  38.709 (7.827)   &  4.851 (0.767)     & 0.801 (0.096)        \\ 
  &  ESR &   35.054 (5.364)   &  \textbf{4.482 (0.72)}     &0.776 (0.087)        \\ \hline

 &	UNIV	&	46.385 (2.967)	&	10.918 (0.946)	&	2.237 (0.147)	\\
	&	FDR	&	48.564 (2.254)	&	9.351 (1.205)	&	1.659 (0.161)	\\
	&	CV	&	40.722 (5.335)	&	4.829 (0.502)	&	0.784 (0.071)		\\
2048&SURE&	46.385 (2.967)&	5.922 (2.612) & 0.805 (0.082)	\\

   &  ABE &   106.813 (16.849)  &   5.427 (0.71)      & 0.692 (0.08)        \\ 
	&	BAMS	&	935.851 (35.811)	&	9.381 (1.068)	&	\textbf{0.453 (0.055)}	\\
   &  SYM BETA &   31.597 (4.606)   &   3.506 (0.508)   &   0.607 (0.091)        \\ 
   &  LOGISTIC &   1223.217 (36.813) &   4.233 (8.3)     & 0.491 (0.063)       \\ 
	&	LPM	&	117.227 (23.502)	&	5.754 (0.925)	&	0.74 (0.106)		\\
   & TPP&   29.453 (2.334)  &  3.31 (0.27)      & 1.959 (0.03)       \\ 
   & $\Gamma$-MINIMAX&   29.527 (5.352)  &   2.922 (0.421)     & 0.491 (0.063)       \\ 
   &  ESR &   \textbf{28.282 (3.997)}  &   \textbf{2.742 (0.382) }   & 0.471 (0.058)      \\ \hline

\end{tabular}
\caption{AMSE and standard deviation (SD) for the shrinkage and thresholding rules considered in the simulation studies. Results for Doppler as underlying function.}\label{tab:doppler}
\end{table}


\begin{table}[H]
\centering
\begin{tabular}{|c|c|c|c|c|}
\hline
 
\textbf{n} & \textbf{Method} & \textbf{SNR=0.2} & \textbf{SNR=1} & \textbf{SNR=3}   \\ \hline \hline
  &        & \textbf{AMSE (SD)} & \textbf{AMSE (SD)} & \textbf{AMSE (SD)} \\ \hline
 &	UNIV	&	44.007 (9.902)	&	7.668 (1.416)	&	1.709 (0.228)	\\
	&	FDR	&	49.896 (5.84)	&	 9.021 (2.992)	&	1.964 (0.877)		\\
	&	CV	&	38.972 (11.893)	&	4.22 (0.943)	&	0.84 (0.129)	\\
512	&SURE&	44.007 (9.902)	& 7.501 (1.715)	&	0.889 (0.152)		\\

   &  ABE     &   107.887 (34.047)  &   5.297 (1.281)    & 0.824 (0.148)       \\ 
	&BAMS	   &970.231 (74.209)	&	11.421 (2.112)	 &	0.537 (0.089)		\\
   & SYM BETA &   26.202 (11.11)  &  2.909 (0.958)  & 0.732 (0.15)  \\ 
   & LOGISTIC&   1221.526 (75.943)  &   7.517 (16.415)  & 0.575 (0.091)        \\ 
	&	LPM	    &  75.794 (39.467)	    &	4.4 (1.574)  &	0.784 (0.17)		\\
   &TPP&  \textbf{ 21.015 (10.484)} & 6.197 (1.141)  & 0.751 (0.124)  \\ 
   & $\Gamma$-MINIMAX&  27.386 (13.18)   & 2.127 (0.705)  & 0.555 (0.092)   \\ 
   &  ESR   & 22.431 (9.499)  &  \textbf{1.829 (0.577)}  & \textbf{0.504 (0.081)}\\ \hline

 &	UNIV	&	34.768 (8.376)&	5.318 (0.777)	&	1.16 (0.125)	\\
	&	FDR	&	47.087 (7.398)	&	6.238 (1.478)	&	1.227 (0.907)		\\
	&	CV	&	27.787 (8.234)	&	2.814 (0.506)	&	0.595 (0.074)	\\
1024&SURE& 34.768 (8.376)	& 4.633 (1.446)&	0.623 (0.087)	\\

   &  ABE &   95.418 (23.751)   &   4.336 (0.918)    & 0.669 (0.11)     \\ 
	&	BAMS	&952.975 (48.514)	& 9.579 (1.461)	&	0.374 (0.057)		\\
   &  SYM BETA &  16.843 (6.423) &   1.665 (0.41) & 0.504 (0.08)       \\ 
   &  LOGISTIC &  1224.953 (51.927) &  4.028 (10.954)& 0.405 (0.06) \\ 
	&	LPM	&	64.434 (27.919)	&	3.165 (1.043)&	0.598 (0.121)\\
   & TPP&   \textbf{10.392 (4.773)}  & 3.098 (0.536)& 0.396 (0.058)  \\ 
   & $\Gamma$-MINIMAX&  17.828 (7.398) & 1.282 (0.333) & 0.387 (0.059)\\ 
   &  ESR  &   13.691 (5.341)  &   \textbf{1.107 (0.249)} &  \textbf{0.352 (0.05)} \\ \hline

 &	UNIV	&	24.868 (5.129) &	3.377 (0.515)	&	0.803 (0.076)	\\
	&	FDR	&	40.171 (11.06)	&	3.974 (1.139)	&	0.762 (0.113)	\\
	&	CV	&	18.565 (4.463)	&	1.9 (0.317)	&	0.424 (0.044)		\\
2048&SURE&24.868 (5.129)	&	2.106 (0.593) & 0.44 (0.053)\\

   &  ABE &    89.589 (15.713)  &   4.08 (0.662)    & 0.571 (0.079)    \\ 
	&	BAMS	&	932.469 (35.223)	&8.556 (0.963)	&	0.262 (0.033)	\\
   &  SYM BETA &   10.895 (3.322)  &  1.118 (0.222)    & 0.32 (0.042)   \\ 
   &  LOGISTIC &  1224.519 (37.127)   &  2.077 (7.556)  & 0.278 (0.036)  \\ 
	&	LPM	&	58.309 (17.79)	&	2.607 (0.652)	&	0.46 (0.073)		\\
   &  TPP&  \textbf{5.521 (2.564)}  &  1.577 (0.259) & \textbf{0.228 (0.028) } \\ 
   & $\Gamma$-MINIMAX &  12.467 (4.72)  &   0.976 (0.2) & 0.268 (0.035) \\ 
   &  ESR  &  8.652 (2.538)&    \textbf{0.825 (0.133)} & 0.241 (0.03)  \\ \hline

\end{tabular}
\caption{AMSE and standard deviation (SD) for the shrinkage and thresholding rules considered in the simulation studies. Results for Heavisine as underlying function.}\label{tab:heavi}
\end{table}


\begin{table}[H]
\centering
\begin{tabular}{|c|c|c|c|c|c|}
\hline
\textbf{Function} &\textbf{Hiperp.}&$\boldsymbol{J_0}$ & \textbf{TPP} & \textbf{SYM BETA} &\textbf{ESR }  \\ \hline \hline
 &  & 0 & 51.154 & 37.072 & \textbf{34.301} \\
Blocks & $\gamma=2, l=1$&1& 42.701 & 34.158  & \textbf{28.624}\\
 & & 3 & 33.485 & \textbf{27.578} & 31.776 \\\hline

 & &1& 50.917 & \textbf{44.974} & 47.158 \\
&& 3 & 29.436& \textbf{29.286} & 29.436  \\
  &$\gamma=2.4, l=5.8$  & 4 & 33.220 & 33.084 & \textbf{32.793} \\
  & & 5 & 45.714 & 45.781 & \textbf{45.596}  \\
 & & 6 & \textbf{79.766} & 79.988  & 79.866 \\\hline \hline

 &  & 0 &50.849& 16.936&  \textbf{13.820}  \\
Heavisine & $\gamma=2, l=1$&1&  11.211 &  15.956 &  \textbf{10.787}\\
 & & 3 &  19.728&\textbf{12.868}&  21.744 \\\hline

 & &1& 50.541 &  \textbf{21.296} &  30.296\\
&& 3 & \textbf{10.394}&  10.547&  10.434 \\
  &$\gamma=2.4, l=5.8$  & 4 & \textbf{19.773}&  19.925&  19.852 \\
  & & 5 & \textbf{38.374}&38.612&38.517 \\
 & & 6 & \textbf{76.609}& 76.938 & 76.842 \\\hline

\end{tabular}
\caption{AMSE of Three-point-prior (TPP), Beta (SYM BETA) and Epanechnikov (ESR) shrinkage rules for $n = 1024$, $\mathrm{SNR} = 0.2$ and Blocks and Heavisine underlying functions according to different primary resolution levels ($J_0$) and choices of the hiperparameters $\gamma$ and $l$ in \eqref{alpha} to obtain $\alpha(j)$.}
\label{tab:alphajo}
\end{table}

\section{Real data application: EEG dataset }

Epilepsy is a neurological disorder characterized by the occurrence of seizures due to excessive or synchronized neuronal activity in the brain. These seizures have several negative consequences, such as the risk of falls, injuries, psychiatric problems, cognitive deficits, and difficulties in achieving academic and professional goals, in addition to increasing the risk of death. In this context, a central question is whether it is possible to predict or detect seizure episodes (preictal state), allowing for actions to neutralize or minimize the impact of the seizures.

We considered EEG data collected by the Neurology and Neurophysiology Unit of the University of Siena, Italy, during a regional research project aimed at describing, detecting, and predicting epileptic seizures and taking appropriate measures to ensure the safety of people. The data are available at \url{https://physionet.org/content/siena-scalp-eeg/1.0.0/}. The dataset was obtained using EB Neuro and Natus Quantum LTM amplifiers, along with reusable silver/gold cup electrodes, which consisted of EEG recordings from 14 epileptic patients monitored with Video-EEG at 512 Hz. The electrodes were arranged according to the international 10-20 System. For more details, see the original study publication \cite{Detti-2020}.

\begin{figure}
    \centering
    \includegraphics[width=.8\linewidth]{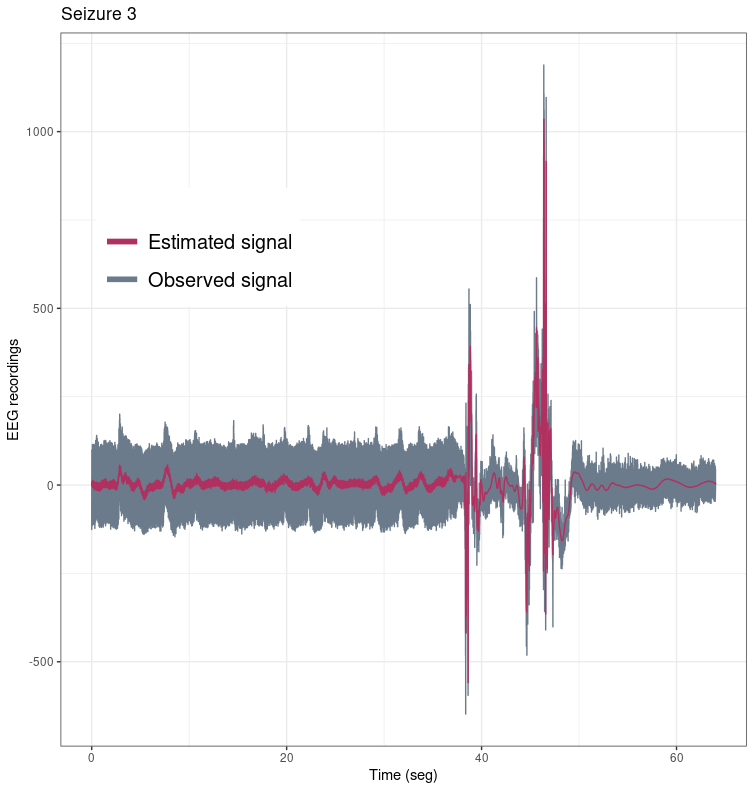} 
    \caption{EEG dataset and the estimated signal using the ESR.}
    \label{eeg1}
\end{figure}

We selected the EEG channel Fp1, with a sample size of $n=2^{15}=32,768$ observations, as shown in Figure \ref{eeg1} in grey. Next, we applied the ESR in order to estimate the signal, which is shown in Figure \ref{eeg1} in red. We considered a Daubechies basis with ten null moments (Daub10) for the DWT and the hyperparameters were chosen according to \eqref{alpha}, \eqref{beta} and \eqref{lambda}, resulting in $\lambda = 0.0006$ and for the highest resolution levels, $\beta(12) = 651.35$, $\beta(13) = 385.99$ and $\beta(14) = 204.07$. The estimated SNR was equal to $1.23$, indicating a high presence of noise in the data. Thus, the application of the ESR is suitable to estimate the signal behind the noisy data. 

\begin{figure}
    \centering
    \includegraphics[width=1\linewidth]{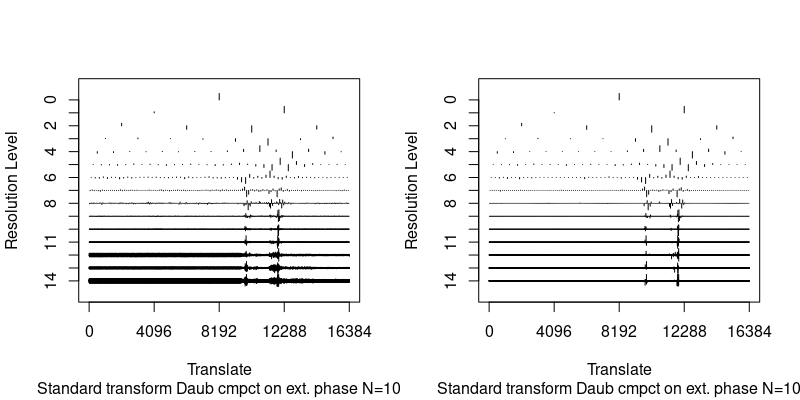} 
    \caption{Empirical (left panel) and estimated (right panel) wavelet coefficients of the EEG dataset.}
    \label{eeg2}
\end{figure}

The denoising action of the rule can also be visualized in Figure \ref{eeg2}, which shows the magnitudes of the empirical wavelet coefficients (left panel) and the estimated (shrunk) wavelet coefficients (right panel) by resolution level. Note that most of the noisy empirical coefficients of the highest resolution levels were severely shrunk around zero by the ESR, leaving only the significant coefficients in the lowest resolution levels.

\section{Final considerations}

This work proposes a wavelet shrinkage rule based on a mixture of a point mass function at zero and the Epanechnikov distribution as a prior for the wavelet coefficients. 

The proposed prior showed to be suitable for modeling wavelet coefficients, combining computational simplicity with desirable statistical properties such as symmetry, unimodality, and flexibility. Moreover, incorporating a suitable prior allows the incorporation of prior knowledge about the coefficients in the estimation procedure, such as sparsity and their support, if they are bounded. 

The associated shrinkage rule had satisfactory performance in the simulation studies, outperforming standard and Bayesian methods in several scenarios with low signal-to-noise ratios. Furthermore, in scenarios where the proposed rule did not perform the best, it was competitive with the best method. The illustration with the EEG dataset reinforced the practical relevance, highlighting the applicability of the rule in real data with high presence of noise. 

Nevertheless, there is room for future developments, such as extending the model to multivariate data or time series, as well as exploring other prior distributions and more efficient hyperparameter selection methods.

\bibliographystyle{plainnat}
\bibliography{references}

\appendix
\clearpage
\section{Proof of the Proposition 1}
Considering that $d|\theta,\sigma^2\sim N(\theta,\sigma^2)$ and $\sigma^2\sim \exp(\lambda)$, $\beta>0$, $\lambda>0$, we have 
\begin{align}\label{ap1}
    \pi(d|\theta) & = \int_{0}^{\infty}\pi(d|\theta, \sigma^2)\pi(\sigma^2|\lambda)d\sigma^2 \nonumber \\
    & = \int_{0}^{\infty} \frac{1}{\sqrt{2\pi}}\frac{1}{\left( \sigma^2 \right)^{1/2}}\exp\left\{ -\frac{(d-\theta)^2}{2\sigma^2} \right\} \lambda \exp\{-\lambda\sigma^2\}d\sigma^2 \nonumber \\
    & = \frac{\lambda}{\sqrt{2\pi}}\int_{0}^{\infty} x^{v -1}\exp\left\{  -Px -\frac{Q}{x} \right\} dx,  
\end{align}
where $Q=\frac{1}{2}(d-\theta)^2$, $P=\lambda$ and $v=\frac{1}{2}$. The last integral can be calculated as
\begin{equation*}
    \int_{0}^{\infty} x^{v -1}\exp\left\{  -Px -\frac{Q}{x} \right\}\, dx =2\left( \frac{Q}{P}\right)^{v/2}K_v\left(2\sqrt{PQ} \right),
\end{equation*}

\noindent where $K_v(x)$ is the second type modified Bessel function with order $v$, which is given by 
\begin{equation*}
    K_v(x)=\frac{\pi}{2}\frac{I_{-v}(x)-I_{v}(x)}{\sin(v\pi)},
\end{equation*}

\noindent where $I_{v}(x)=\sum_{k=0}^{\infty}\frac{1}{k!(k+v)!}\left( \frac{x}{2}\right)^{2k+v}$. For $v=\frac{1}{2}$, there are two expressions for $I_{v}(x)$, 
\begin{equation*}
    I_{1/2}(x)=\sqrt{\frac{2}{\pi x}}\sinh(x)\quad  \mathrm{and} \quad I_{-1/2}(x)=\sqrt{\frac{2}{\pi x}}\cosh(x),
\end{equation*}

\noindent then, we have that  $K_{1/2}(x)=\sqrt{\frac{2}{\pi x}}e^{-x}$ and
\begin{equation}\label{ap2}
\int_{0}^{\infty} x^{v -1}\exp\left\{  -Px -\frac{Q}{x} \right\}\, dx=\sqrt{\pi/P}e^{-2\sqrt{PQ}}. 
\end{equation}
Substituting \eqref{ap2} in \eqref{ap1},
\begin{align*}
    \pi(d|\theta) &=\frac{\lambda}{\sqrt{2\pi}}\sqrt{\pi/P}e^{-2\sqrt{PQ}}\\
    & =\frac{\sqrt{2\lambda}}{2}e^{-2\sqrt{\lambda\frac{1}{2}(d-\theta)^2}}\\
    & =\frac{\sqrt{2\lambda}}{2}e^{-\sqrt{2\lambda}|d-\theta|},
\end{align*}
\noindent i.e., $\quad d|\theta\sim \mathcal{ED}\left(\theta,\frac{1}{ \sqrt{2\lambda}}\right)$. 

On the other hand, the marginal distribution of $d$ is 
\begin{align}\label{ap3}
    m(d) & =\int_{-\beta}^{\beta}\pi(d|\theta)\pi(\theta)d\theta = \int_{-\beta}^{\beta}\frac{\sqrt{2\lambda}}{2}e^{-\sqrt{2\lambda}|d-\theta|} \frac{3}{4\beta^3}\left(\beta^2 -\theta^2\right)d\theta \nonumber \\
& = \frac{3\sqrt{2\lambda}}{8\beta^3}\int_{-\beta}^{\beta}\left(\beta^2 -\theta^2\right)e^{-\sqrt{2\lambda}|d-\theta|}d\theta.
\end{align}

\noindent Solving $I_1=\int_{-\beta}^{\beta}\left(\beta^2 -\theta^2\right)e^{-\sqrt{2\lambda}|d-\theta|}d\theta $, we have that

\begin{align}\label{ap4}
I_1 & = \int_{-\beta}^{d}\left(\beta^2 -\theta^2\right)e^{-\sqrt{2\lambda}(d-\theta)}d\theta + \int_{d}^{\beta}\left(\beta^2 -\theta^2\right)e^{-\sqrt{2\lambda}(\theta-d)}d\theta \nonumber\\
& = \int_{-\beta}^{d}\beta^2 e^{-\sqrt{2\lambda}(d-\theta)}d\theta-\int_{d}^{\beta}\theta e^{-\sqrt{2\lambda}(d-\theta)}d\theta+ \int_{d}^{\beta}\beta^2 e^{-\sqrt{2\lambda}(\theta-d)}d\theta \nonumber\\
& -\int_{d}^{\beta}\theta^2 e^{-\sqrt{2\lambda}(\theta-d)}d\theta \nonumber\\
& =\beta^2 e^{-\sqrt{2\lambda}d} \int_{-\beta}^{d}e^{\sqrt{2\lambda}\theta}d\theta-e^{-\sqrt{2\lambda}d} \int_{d}^{\beta}\theta^2 e^{\sqrt{2\lambda}\theta}d\theta+ \beta^2 e^{\sqrt{2\lambda}d}\int_{d}^{\beta} e^{-\sqrt{2\lambda}\theta}d\theta \nonumber\\
& -e^{\sqrt{2\lambda}d} \int_{d}^{\beta}\theta^2 e^{-\sqrt{2\lambda}\theta}d\theta \nonumber\\
& = \beta^2 e^{-\sqrt{2\lambda}d}\left( \frac{e^{\sqrt{2\lambda}\theta }}{\sqrt{2\lambda}}\right)\bigg|_{-\beta}^{d} - e^{-\sqrt{2\lambda}d}\left(\frac{\theta^2 e^{\sqrt{2\lambda}\theta}}{\sqrt{2\lambda}} -\frac{\theta e^{\sqrt{2\lambda}\theta}}{\lambda} + \frac{ e^{\sqrt{2\lambda}\theta}}{\lambda\sqrt{2\lambda}}\right)\bigg|_{-\beta}^{d} \nonumber \\
& + \beta^2 e^{\sqrt{2\lambda}d}\left( \frac{e^{-\sqrt{2\lambda}\theta }}{-\sqrt{2\lambda}}\right)\bigg|_{d}^{\beta} - e^{\sqrt{2\lambda}d}\left(\frac{\theta^2 e^{\sqrt{2\lambda}\theta}}{-\sqrt{2\lambda}} -\frac{\theta e^{-\sqrt{2\lambda}\theta}}{\lambda} + \frac{ e^{-\sqrt{2\lambda}\theta}}{-\lambda\sqrt{2\lambda}}\right)\bigg|_{d}^{\beta} \nonumber\\
& = \frac{\beta^2}{\sqrt{2\lambda}} - \frac{\beta^2 e^{-\sqrt{2\lambda}(\beta+d)}}{\sqrt{2\lambda}} - e^{-\sqrt{2\lambda}d} \left(\frac{d^2 e^{\sqrt{2\lambda}d}}{\sqrt{2\lambda}} - \frac{d e^{\sqrt{2\lambda}d}}{\lambda} + \frac{e^{\sqrt{2\lambda}d}}{\lambda\sqrt{2\lambda}} - \frac{\beta^2 e^{-\sqrt{2\lambda}\beta}}{\sqrt{2\lambda}}\right. \nonumber\\
& \left. - \frac{\beta e^{-\sqrt{2\lambda}\beta}}{\lambda} - \frac{e^{-\sqrt{2\lambda}\beta}}{\lambda\sqrt{2\lambda}}\right)  - \frac{\beta^2 e^{-\sqrt{2\lambda}(\beta-d)}}{\sqrt{2\lambda}} + \frac{\beta^2}{\sqrt{2\lambda}} - e^{\sqrt{2\lambda}d}\left( -\frac{\beta^2 e^{-\sqrt{2\lambda}\beta}}{\sqrt{2\lambda}} - \frac{\beta e^{-\sqrt{2\lambda}\beta}}{\lambda} \right. \nonumber\\
& \left. - \frac{e^{-\sqrt{2\lambda}\beta}}{\lambda\sqrt{2\lambda}} +\frac{d^2 e^{-\sqrt{2\lambda}d} }{\sqrt{2\lambda}} + \frac{d e^{-\sqrt{2\lambda}d}}{\lambda}+ \frac{e^{-\sqrt{2\lambda}d}}{\lambda\sqrt{2\lambda}}\right) \nonumber\\
& = \frac{\beta^2}{\sqrt{2\lambda}} - \frac{\beta^2 e^{-\sqrt{2\lambda}(\beta+d) }}{\sqrt{2\lambda}} - \frac{d^2}{\sqrt{2\lambda}}+ \frac{d}{\lambda} - \frac{1}{\lambda\sqrt{2\lambda}}+ \frac{\beta^2 e^{-\sqrt{2\lambda}(\beta+d)}}{\sqrt{2\lambda}} + \frac{\beta e^{-\sqrt{2\lambda}(\beta+d)}}{\lambda} \nonumber\\
& +  \frac{e^{-\sqrt{2\lambda}(\beta+d)}}{\lambda\sqrt{2\lambda}} - \frac{\beta^2 e^{-\sqrt{2\lambda}(\beta-d) }}{\sqrt{2\lambda}} +  \frac{\beta^2}{\sqrt{2\lambda}} + \frac{\beta^2 e^{-\sqrt{2\lambda}(\beta-d) }}{\sqrt{2\lambda}} + \frac{\beta e^{-\sqrt{2\lambda}(\beta-d) }}{\lambda} \nonumber\\
& + \frac{e^{-\sqrt{2\lambda}(\beta-d) }}{\lambda\sqrt{2\lambda}} - \frac{d^2}{\sqrt{2\lambda}} - \frac{d}{\lambda} - \frac{1}{\lambda\sqrt{2\lambda}} \nonumber\\
& = \frac{\beta}{\lambda}\left( e^{-\sqrt{2\lambda}(\beta+d)} + e^{-\sqrt{2\lambda}(\beta-d)} \right) + \frac{2\beta^2}{\sqrt{2\lambda}} - \frac{2}{\sqrt{2\lambda}}\left( d^2 + \frac{1}{\lambda}\right) \nonumber\\
& = \frac{\beta}{\lambda}\left( e^{-\sqrt{2\lambda}(\beta+d)} + e^{-\sqrt{2\lambda}(\beta-d)} \right) + \frac{2}{\sqrt{2\lambda}}\left( \beta^2 - d^2 -\frac{1}{\lambda} \right).
\end{align}

Substituting \eqref{ap4} in \eqref{ap3}, it follows that
\begin{equation} \label{ap5}
    m(d)=\frac{3\sqrt{2\lambda}}{8\beta^3}\left[ \frac{\beta}{\lambda}\left( e^{-\sqrt{2\lambda}(\beta+d) + e^{-\sqrt{2\lambda}(\beta-d) } }\right)+ \frac{2}{\sqrt{2\lambda}}\left( \beta^2 - d^2 -\frac{1}{\lambda}\right)\right].
\end{equation}
Under the squared error loss function, the Bayes rule $\delta(d)$ is the posterior mean of $\theta|d$, i.e,
\begin{align}\label{ap6}
 \delta(d) = E(\theta|d) & =\int_{\theta\in\Theta}\theta\pi(\theta|d)d\theta= \int_{\theta\in\Theta}\theta \frac{\pi(d|\theta)\pi(\theta)}{m(d)}d\theta \nonumber\\
& = \frac{1}{m(d)}\int_{-\beta}^{\beta}\theta\frac{\sqrt{2\lambda}}{2}e^{-\sqrt{2\lambda}|d-\theta|}\frac{3}{4\beta^3}\left(\beta^2 -\theta^2\right)d\theta \nonumber\\
& = \frac{3\sqrt{ 2 \lambda } }{ 8\beta^3 m(d)}\int_{-\beta}^{\beta}\theta (\beta^2 - \theta^2 )e^{-\sqrt{2\lambda}|d-\theta|}d\theta.
\end{align}
\noindent Let $I_2=\int_{-\beta}^{\beta}\theta (\beta^2 - \theta^2 )e^{-\sqrt{2\lambda}|d-\theta|}d\theta$. Then
\begin{align}\label{ap7}
I_2 & = \int_{-\beta}^{\beta}\beta^2 \theta  e^{-\sqrt{2\lambda}|d-\theta|}d\theta - \int_{-\beta}^{\beta}\theta^3 e^{-\sqrt{2\lambda}|d-\theta|}d\theta \nonumber\\
& =  \int_{-\beta}^{d}\beta^2\theta e^{-\sqrt{2\lambda}(d-\theta)}d\theta + \int_{d}^{\beta}\beta^2 \theta e^{-\sqrt{2\lambda}(\theta-d)}d\theta \nonumber\\
& -  \int_{-\beta}^{d}\theta^3 e^{-\sqrt{2\lambda}(d-\theta)}d\theta -  \int_{d}^{\beta}\theta^3 e^{-\sqrt{2\lambda}(\theta-d)}d\theta \nonumber\\
& = \beta^2 e^{-\sqrt{2\lambda}d} \int_{-\beta}^{d}\theta e^{\sqrt{2\lambda}\theta}d\theta +\beta^2 e^{\sqrt{2\lambda}d}  \int_{-\beta}^{d}\theta e^{-\sqrt{2\lambda}\theta}d\theta \nonumber\\
& - e^{-\sqrt{2\lambda}d} \int_{-\beta}^{d}\theta^3 e^{\sqrt{2\lambda}\theta}d\theta - e^{\sqrt{2\lambda}d} \int_{d}^{\beta}\theta^3 e^{-\sqrt{2\lambda}\theta}d\theta.
\end{align}

To solve \eqref{ap7}, we use the following results:
\begin{itemize}
    \item $\int \theta e^{a \theta}d\theta = \frac{e^{a \theta}}{a^2}+C$
    
    \item $\int \theta^3 e^{a \theta}d\theta = \frac{e^{a \theta}\left(a^3 \theta^3 - 3 a^2 \theta^2 + 6 a\theta - 6 \right) }{a^4}+C$
\end{itemize}

Then, we solve each integral of \eqref{ap7} as

\begin{itemize}
    \item[] $R1=\beta^2 e^{-\sqrt{2\lambda}d}\int_{-\beta}^{d}\theta e^{\sqrt{2\lambda}\theta}d\theta= \frac{\beta^2 \left(\sqrt{2\lambda}d-1 \right)}{2\lambda}+ \frac{\beta^2 \left(\sqrt{2\lambda}d+1 \right)e^{-\sqrt{2\lambda}(\beta+d)} }{2\lambda}$
    
    \item[] $R2=\beta^2 e^{\sqrt{2\lambda}d}\int_{d}^{\beta}\theta e^{-\sqrt{2\lambda}\theta}d\theta= \frac{\beta^2 \left(\sqrt{2\lambda}d+1 \right)}{2\lambda}- \frac{\beta^2 \left(\sqrt{2\lambda}\beta+1 \right)e^{-\sqrt{2\lambda}(\beta-d)} }{2\lambda}$

    \item[] $R3=e^{-\sqrt{2\lambda}d}\int_{-\beta}^{d}\theta^3 e^{\sqrt{2\lambda}\theta}d\theta=\frac{\lambda\sqrt{2\lambda}d^3 - 3\lambda d^2 + 3 \sqrt{2\lambda}d-3 }{2\lambda^2}+\varLambda e^{-\sqrt{2\lambda}(\beta+d)} $

    \item[] $R4=e^{\sqrt{2\lambda}d}\int_{d}^{\beta}\theta^3 e^{-\sqrt{2\lambda}\theta}d\theta=\frac{\lambda\sqrt{2\lambda}d^3 + 3\lambda d^2 + 3 \sqrt{2\lambda}d+3 }{2\lambda^2} - \varLambda e^{-\sqrt{2\lambda}(\beta-d)} $,
    
\end{itemize}

\noindent where $\varLambda=\frac{\lambda\sqrt{2\lambda}\beta^3 + 3\lambda \beta^2 + 3 \sqrt{2\lambda}\beta +3 }{2\lambda^2}$. Substituting $R1-R4$ in \eqref{ap7}, we have that
\begin{align} \label{ap8}
    I_2 & = R1+R2-(R3+R4) \nonumber\\
    & = \frac{2\lambda\beta^2 +3\sqrt{2\lambda}\beta+3}{2\lambda^2}\Delta_1+ \frac{(\lambda\beta^2 -3)\sqrt{2\lambda}d -\lambda\sqrt{2\lambda}d^3}{\lambda^2},
\end{align}
\noindent where $\Delta_1=e^{-\sqrt{2\lambda}(\beta-d)}- e^{-\sqrt{2\lambda}(\beta+d)}$. Thus, substituting \eqref{ap5} and \eqref{ap8} in \eqref{ap6},
\begin{equation}
\label{ap9}
    \delta(d)=\frac{\frac{2\lambda\beta^2 +3\sqrt{2\lambda}\beta+3}{2\lambda^2}\Delta_1 + \frac{(\lambda\beta^2 -3)\sqrt{2\lambda}d -\lambda\sqrt{2\lambda}d^3}{\lambda^2}}{\frac{\beta}{\lambda} \Delta_2  + \frac{2}{\sqrt{2\lambda}}\left( \beta^2 - d^2 -\frac{1}{\lambda}\right)},
\end{equation}
\noindent where $\Delta_2= e^{-\sqrt{2\lambda}(\beta+d)}+ e^{-\sqrt{2\lambda}(\beta-d) } $.

Finally, considering the prior distribution \eqref{prior} and 
and $\boldsymbol{\mathit{L}}(d|\theta)$ the likelihood function, the posterior distribution of $\theta|d$ is given by
\begin{align*}
    \pi^{*}(\theta|d) & =\frac{\boldsymbol{\mathit{L}}(d|\theta) \pi ^{*}(\theta)}{\int_{\theta \in \Theta}\boldsymbol{\mathit{L}}(d|\theta)\pi^{*}(\theta)d\theta}= \frac{ ( \alpha\delta_0 +(1-\alpha)g(\theta;\beta))\boldsymbol{\mathit{L}}(d|\theta) }{\int_{\theta\in\Theta} (\alpha\delta_0 +(1-\alpha)g(\theta;\beta))\boldsymbol{\mathit{L}}(d|\theta) d\theta },
\end{align*}
\noindent where $\alpha\in (0,1)$. Thus, the proposed Epanechnikov shrinkage rule is given by
\begin{align*}
\delta^{*}(d) & = E(\theta^{*}|d) \\
& = \frac{(1-\alpha)\int_{-\beta}^{\beta}\theta g(\theta;\beta)L(d|\theta)d\theta}{\alpha \boldsymbol{\mathit{L}}(d|\theta=0)+(1-\alpha) \int_{-\beta}^{\beta}g(\theta;\beta)\boldsymbol{\mathit{L}}(d|\theta)d\theta}\\
& = \frac{(1-\alpha)\int_{-\beta}^{\beta}\theta g(\theta;\beta)L(d|\theta)d\theta}{\alpha \boldsymbol{\mathit{L}}(d|\theta=0)+(1-\alpha) \int_{-\beta}^{\beta}g(\theta;\beta)\boldsymbol{\mathit{L}}(d|\theta)d\theta}\\
& = \frac{(1-\alpha)\int_{-\beta}^{\beta}\theta \frac{3}{4\beta^3}(\beta^2 -\theta^2)\frac{\sqrt{2\lambda}}{2}e^{-\sqrt{2\lambda}|d-\theta|}d\theta}{\alpha \frac{\sqrt{2\lambda}}{2}e^{-\sqrt{2\lambda}|d|}+(1-\alpha) \int_{-\beta}^{\beta}\frac{3}{4\beta^3}(\beta^2 -\theta^2)\frac{\sqrt{2\lambda}}{2}e^{-\sqrt{2\lambda}|d-\theta|}d\theta} \\
& = \frac{(1-\alpha)m(d)\delta(d)}{\alpha \mathcal{ED}(0,\frac{1}{\sqrt{2\lambda}})+ (1-\alpha)m(d)},
\end{align*}

\noindent where $m(d)$ and $\delta(d)$ are given by \eqref{ap5} and \eqref{ap9} respectively.  $\qed$.

\end{document}